\documentclass[useAMS,usenatbib]{mn2e}

\usepackage{graphicx}

\usepackage{aas_macros}
\usepackage{rotating}
\def\Mo{M_\odot} 

\def\apjs{ApJS}
\def\lsim{\mathrel{\rlap{\lower4pt\hbox{\hskip1pt$\sim$}}
    \raise1pt\hbox{$<$}}}                % less than or approx. symbol
\def\gsim{\mathrel{\rlap{\lower4pt\hbox{\hskip1pt$\sim$}}
    \raise1pt\hbox{$>$}}}                % greater than or approx. symbol
\begin{document}

\title{The host galaxies of core-collapse supernovae and gamma ray bursts}
\author[K.M. Svensson et al]{K. M. Svensson$^{1}$, A. J. Levan$^{1}$, N. R. Tanvir$^{2}$, A. S. Fruchter$^3$  L.-G. Strolger$^4$\\
$^{1}$ Department of Physics, University of Warwick, Coventry CV4 7AL, UK \\
$^{2}$ Department of Physics and Astronomy, University of Leicester, University Road, Leicester, LE1 7RH, UK\\
$^{3}$ Space Telescope Science Institute, 3700 San Martin Drive, Baltimore, MD 21218, USA \\
$^{4}$ Dept. of Physics and Astronomy, Western Kentucky University,1906 College Heights Blvd 11077,Bowling Green, KY 42101-1077,USA}

\date{Received;Accepted}

\pagerange{\pageref{firstpage}--\pageref{lastpage}} \pubyear{2010}

\maketitle

\label{firstpage}

\begin{abstract}
  We present a comparative study of the galactic and small scale environments of gamma-ray bursts (GRB)
  and core collapse supernovae (CCSN). We use a sample of 34 GRB hosts at $z<1.2$, 
  and a comparison sample of 58 supernova hosts located within
  the Great Observatories Origins Deep Survey footprint. We fit template spectra to
  the available photometric data, which span the range  0.45-24$\mu$m, and extract
  absolute magnitudes, stellar masses and star formation rates from the resulting fits.
  Our results broadly corroborate previous findings, but offer significant enhancements
  in spectral coverage and a factor 2-3 increase in sample size. 
  Specifically, we find that 
  CCSN occur frequently in massive spirals (spiral fraction $\sim$ 50\%). In contrast GRBs
  occur in small, relatively low mass galaxies with high specific and surface star formation rates,
  and have a spiral fraction of only $\sim 10$\%. A
  comparison of the rest frame absolute magnitudes of the GRB and CCSN sample is less 
  conclusive than found in previous work, suggesting that while GRB hosts are typically both
  smaller and bluer than those of CCSN their total blue light luminosities are only slightly lower. 
  We suggest this is likely due to rapid periods of intensified
  star formation activity, as indicated by the high specific
  star formation rates, which both create the GRB progenitors
  and briefly significantly enhance the host galaxy blue luminosity. Finally, our analysis
  of local environments of GRBs and CCSN shows that GRBs are highly concentrated
  on their host light, and further occur in regions of higher absolute surface luminosity
  than CCSN.
\end{abstract}


% ############################################################################################################
% 
% ############################################################################################################

\section{Introduction}

Core-collapse supernovae (CCSN) mark the end-points in
the lives of short-lived (lifetime $\lsim$ few$\times 10^7$ years), massive
stars (M $\gsim$8M$_{\odot}$).
The selection of galaxies via the presence of a CCSN thus
provides, in principle, an ideal mechanism for the detection of
star forming galaxies at a range of redshifts.
Long duration GRBs are closely related to CCSN, and offer similar advantages as tracers of star formation, which have been widely discussed in e.g.
\cite{2005MNRAS.362..245J,2006A&A...447..897J,1998MNRAS.297L..17M}. 
Specifically, both CCSN and GRB
production requires only a single stellar progenitor, and so they
select galaxies {\em independently} of the galaxy luminosity.
By doing so they can point at galaxies too faint to be included in 
flux limited surveys,
potentially providing a handle on the faint end of the galaxy
luminosity function at high-z. Unlike GRBs however, CCSN are
less affected by metallicity effects, and hence they provide a more complete  
selection of the collapse of stars with
initial main sequence masses in excess of $\sim$8M$_{\odot}$.
Therefore, a census of supernova host galaxies is providing
a census of essentially all massive star formation at a given redshift.

One drawback in the use of supernovae as a direct probe of star
formation has been the inability to pursue searches for CCSN
beyond $z \sim 1$, due to the limitations of current technology. Out to this distance the luminosity
function, and star formation rate are reasonably well constrained 
through
other methods. However, the installation of Wide Field Camera 3 on {\it HST}, and in the longer
term the launch of {\it JWST} offer the opportunity to push this
to much higher redshift. Nonetheless, in the interim period their potential
use to ``calibrate" environmental dependencies in GRBs, and
other star-forming galaxy samples, motivates
their study.

A complication in the use of SN comes from understanding biases in their
observed rate introduced by dust extinction within their hosts. 
While the highly penetrating $\gamma$ and X-ray's from GRBs can largely
circumvent problems with local extinction this is not necessarily the case
for their optical afterglows. CCSN, which are several magnitudes fainter
at peak than a typical GRB optical afterglow \citep[see e.g.][(Figure 9) for an extreme example]{2008arXiv0812.1217T,2009ApJ...691..723B}, are even
more prone to non-detection due to host galaxy extinction. 
In practise, the extent to which 
extinction biases the detection of either GRB optical afterglows or CCSN remains
poorly understood, although it is likely to impact both 
\citep[e.g.][]{2003A&A...401..519M,2006Natur.441..463F,2007ApJ...669.1098R,2006ApJ...647..471L}

Effort has already been invested in studying SN hosts, and the
locations of SN within them. In particular this has focused on large
samples of SN at low redshift, for example those found by, or 
overlapping
with, the Sloan Digital Sky Survey e.g. \cite{2007arXiv0707.0690P}  or those 
found in galaxies targeted by other surveys e.g. \cite{2006A&A...453...57J}.
These surveys offer insight into SN host properties and locations, and using local SN, with small
angular distances, allow the environments to be probed in detail. However, locally
discovered supernovae have historically been found by targeted searches
of specific galaxy catalogues, producing a bias towards brighter host galaxies. More
recent searches (e.g. SDSS and SN Factory, and in the near future Skymapper and Pan-STARRS) 
avoid this by repeatedly tiling blank regions of sky, 
although they typically find more distant SN. Comparisons of these hosts suggest
that while SN globally trace star formation the relative fractions of Ib/c increase
in highly metal  enriched environments, likely reflecting the tendency for massive
stars to loose their hydrogen envelopes via radiatively driven winds at higher metallicity
\citep{2007arXiv0707.0690P}. 

All CCSN, by their nature, indicate the formation of massive stars in their
hosts, while the locations of the supernovae within their hosts can also
be strongly diagnostic. \cite{2006Natur.441..463F}(hereafter F06) used a new pixel statistic 
(essentially the fraction of light contained in regions of lower surface
brightness than the region containing SN or GRB) to show that GRBs are highly concentrated
on the light of their hosts, and likely favour a much more massive and shorter lived progenitor
than CCSN, which trace blue light within their host galaxy. 
Utilising this technique on a lower redshift sample of CCSN found
in the SDSS fields,
\cite{2007arXiv0712.0430K} show that SN Ic are also highly concentrated on the
brightest regions of their hosts, a distribution very similar to GRBs. 
This may suggest that both GRBs and SN Ic originate only from the most
massive stars \citep{2007MNRAS.376.1285L}.  \cite{2006A&A...453...57J} take an 
alternative approach of using H$\alpha$ images and similarly find
that SN Ib/c are more concentrated on their hosts. They suggest that
this may be due to the expulsion of SN II progenitors from their star forming regions
with moderate velocities, rather than an intrinsic tendency for SN Ib/c to
lie on brighter regions of their hosts. Should SNII typically 
originate from less massive stars than SN Ib/c then this may be
expected since the transverse distances travelled over the stellar lifetime
would be larger for less massive (and hence longer lived) stars.

 Although there is a growing consensus that GRBs originate from different environments than the bulk of CCSN,
 it is not yet clear how well the global properties of the whole host galaxy are 
 evidence of this. \cite{2008arXiv0803.2718S} note that global metallicity measurements of
 GRB hosts are predominantly subsolar\footnote{Although at times this conclusion depends on
 an assumption about the ionisation parameter within the host}.
 This agrees with theoretical models of GRB production, which favour lower metallicity 
 environments \citep[e.g.][]{2003ApJ...591..288H}.  
 Furthermore, a study by \cite{2008AJ....135.1136M} suggested that SN Ic not associated with GRBs tend
 to originate from more metal rich environments than SN Ic with a GRB associated. These
 authors  also suggested that sub solar (20 to 60 percent of solar) metallicity is required to produce
 a GRB. A complication of testing this hypothesis is that 
 metallicity can vary by several tenths of a dex within the hosts, both by localised enrichment \citep[e.g. the IFU measurements by][]{2008A&A...490...45C}  and
 due to a radial gradient \citep[e.g.][]{1997ApJ...489...63G,2000A&A...363..537R}. This makes spatially resolved spectroscopy, or direct
 measurements of metallicity from the afterglow spectrum valuable. However,
 this is impossible for a significant fraction of GRBs, since the angular distances are
 too small to resolve the hosts into many resolution elements. Thus, while not an ideal 
 measure, estimates of the stellar mass or luminosity of the hosts can be used as a proxy for metallicity, and when
 averaged over a large number of hosts should still provide robust statements about 
 CCSN and GRB environments. 
 
Here we investigate the multi-wavelength properties of a sample of
CCSN host galaxies observed by the GOODS  (Great Observatories 
Origins Deep Survey),
and PANS (Probing Acceleration Now with Supernovae) surveys,
and compare these to those of GRBs. These galaxies, lying at comparable
redshift to many GRBs, although at distinctly lower-$z$ than the mean value of $\sim2.5$ \citep{2006A&A...447..897J},
offer the opportunity for direct comparison of derived physical properties (e.g. 
mass, star formation rate), without the need to worry
about evolutionary effects in either the galaxy luminosity function, or,
in the case of GRBs, the universal evolution of metallicity. 
Using a large, multi-wavelength (optical through 
mid-IR) dataset we derive physical parameters for the
host galaxies of CCSN and GRBs. This includes, rest frame luminosities, star formation rates, 
stellar mass and
surface brightness at the GRB or SN location. Considering possible bias effects that might be present in both samples,
our results  broadly echo those of previous work that GRB hosts are typically
smaller and less massive than those of CCSN, most likely
due to metallicity bias. GRBs also
originate in brighter locations, consistent with their origin in more massive stars.

% ############################################################################################################
%
% ############################################################################################################

\section{Host galaxy samples}
\subsection{Supernovae in GOODS and PANS}
The GOODS \citep{2004ApJ...600L..93G}
survey undertook observations in two fields, centred on the
Hubble Deep Field North and Chandra Deep Field South.
These observations included deep observations with the
{\em Hubble Space Telescope} using the
Advanced Camera for Surveys (ACS) in the F450W(B), F606W(broad V/R), F814W(I) and 
F850LP(Z)
filters. Rather than obtain the images in a single epoch the observations
were made roughly every 45 days, to be
sensitive to the rise time of SN Ia at $z \sim 1$ \citep[see eg][]{2004ApJ...600L.163R} 
As well as detecting a number of SN Ia, these observations also located
numerous core collapse supernovae (\citep[e.g.][and Dahlen et al. in prep]{2004ApJ...613..200S,2008ApJ...681..462D}
with a mean redshift of $z \sim 0.6
$ (CCSN are generally
less luminous at maximum than SN Ia, and so visible over a smaller 
volume
in a flux limited sample). These SN host galaxies form an excellent sample
for further study, by virtue of their selection in a blind survey, independent of
galaxy luminosity (in contrast to many low-z SN searches which are targeted
at specific galaxy catalogues), and because of the wide range of supporting
data covering the blue optical to mid-IR regions. 

\begin{figure*}
  \includegraphics[scale=0.9]{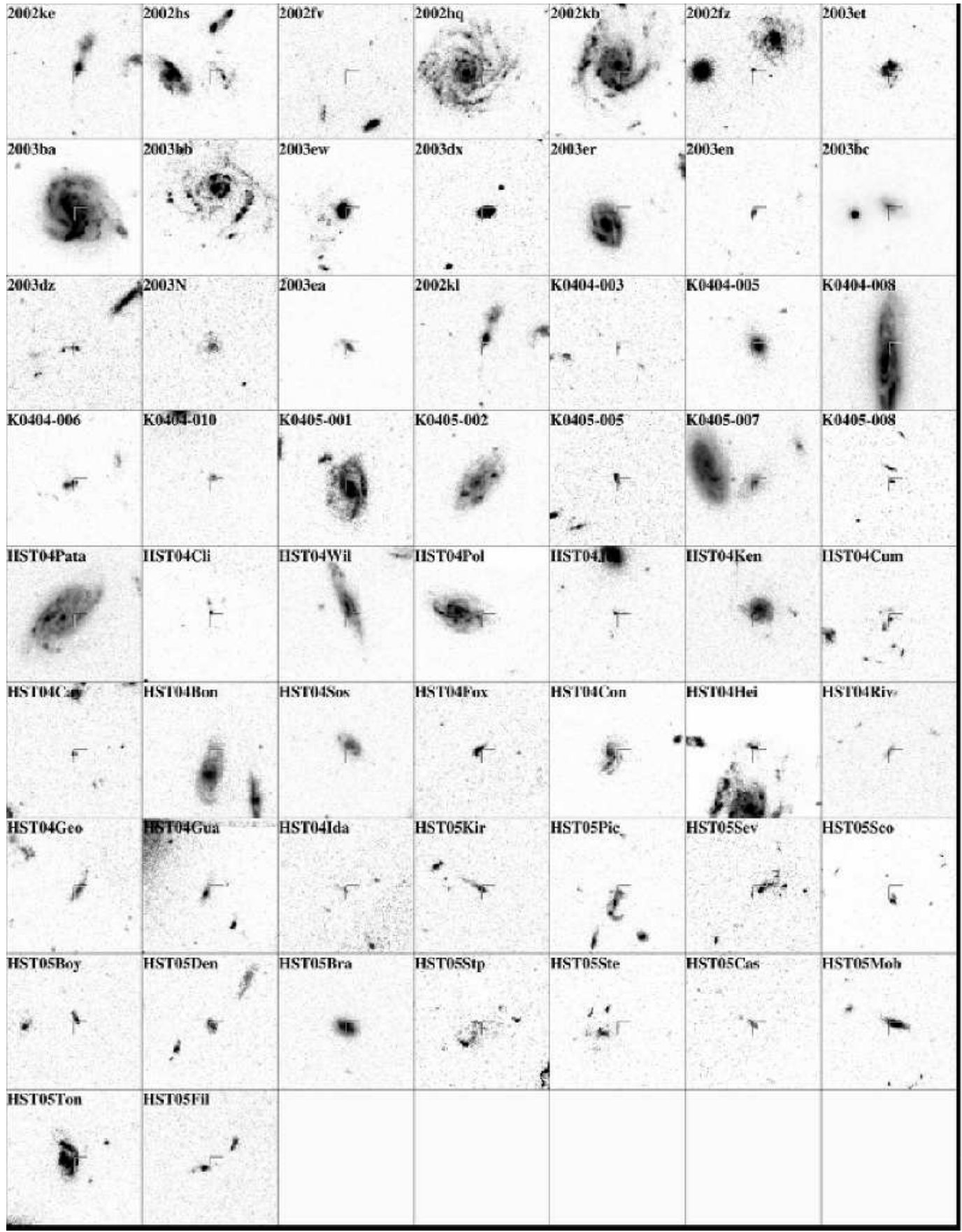}
  \caption{Mosaic image of the 58 CCSN host galaxies in the GOODS fields. These V-band images have a width of 7.5
    arcseconds and the location of the Supernovae on the host is marked with a cross-hair.}
  \label{snmos}
\end{figure*}

These data, in addition to that secured by {\it HST} and described above, encompassed
large programmes with {\em Spitzer} and also a concerted effort from 
ground based observatories to secure complementary near-IR observations
and redshift catalogues. ACS images of the resulting sample of CCSN hosts are
shown in Figure ~\ref{snmos}.

Each SN discovered in GOODS or subsequently PANS is typed based on the available
photometric and spectroscopic data on both the SN and its host galaxy. The means of
this typing is described in \cite{2004ApJ...613..200S}, its outcome is that the confidence in
the typing of a given supernova is given by the assignation of a ``medal". 
These medals, termed Gold, Silver or Bronze reflect both the quality and quantity of
data available to type
the SN. The optimal diagnostic is obviously a spectrum of the SN itself, demonstrating the
clear presence (or absence) of hydrogen. Spectroscopically typed SN are given a Gold medal. 
In the absence of a spectrum the diagnostics used are the lightcurve shape, its peak absolute magnitude, the type of host galaxy and its U-B colour. Initially the lightcurve shape is compared
to that of a SN Ia. If this fit is poor, but the lightcurve well sampled then the transient 
is assigned as a CCSN with a Silver medal.
If the lightcurve is inconclusive, but the
host galaxy appears to be starforming then (in general) the SN is typed as
CCSN with a Bronze medal. Hence, it is possible that the inclusion of Bronze CCSN 
introduces a small number of SN Ia into the CCSN sample. We discuss this issue
, and other selection effects, further in section 8.
For further details on the algorithms for the classification of each SN the reader is referred to 
\cite{2004ApJ...613..200S}.

\begin{figure*}
  \includegraphics[scale=0.85]{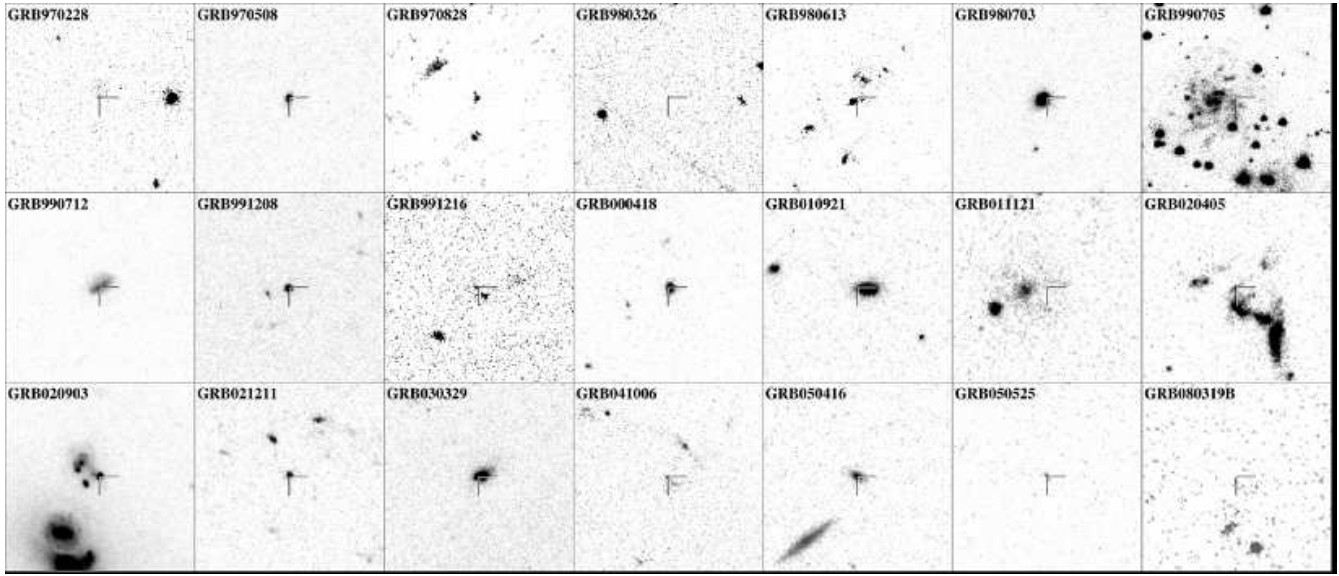}
  \caption{Mosaic image of GRB host galaxies with {\em HST} imaging. The images are 7.5 arcseconds wide, and the locations of the
    GRBs on the host is marked with a cross-hair.}
  \label{grbmos}
\end{figure*}

\begin{figure}
  \includegraphics[scale=0.65]{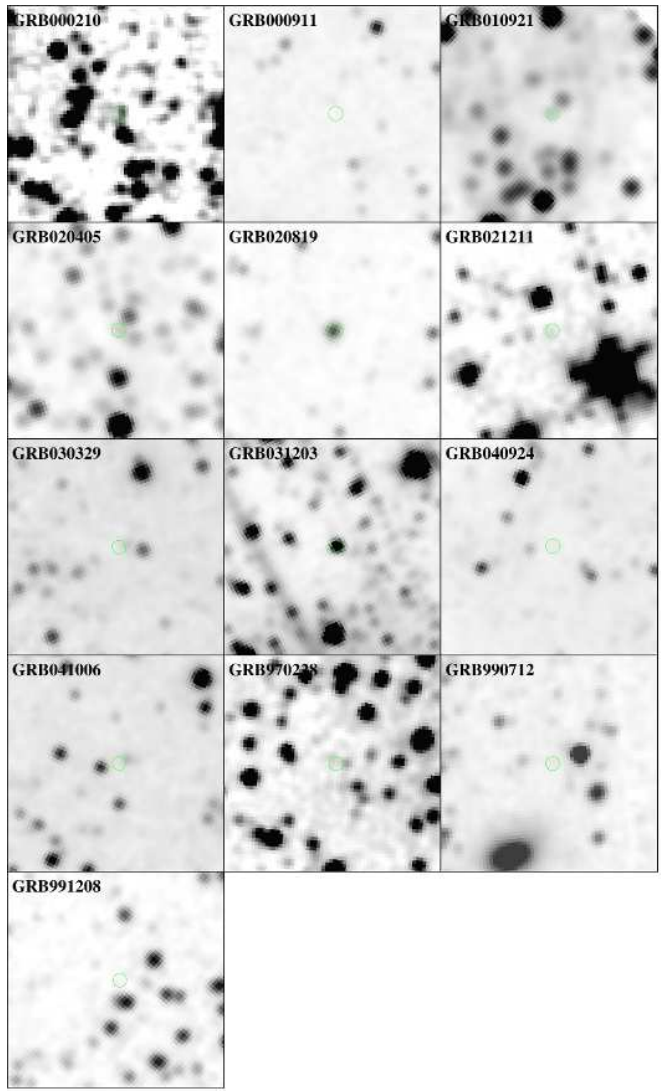}
  \caption{Mosaic image showing the GRB hosts observed with {\em Spitzer} IRAC. Images are in $3.6 \mu m$ where available, otherwise in $4.5 \mu m$.
The width of each tile is $\sim 80$ arcseconds.}
  \label{IRmos}
\end{figure} 
  
\begin{figure}
  \includegraphics[scale=0.4]{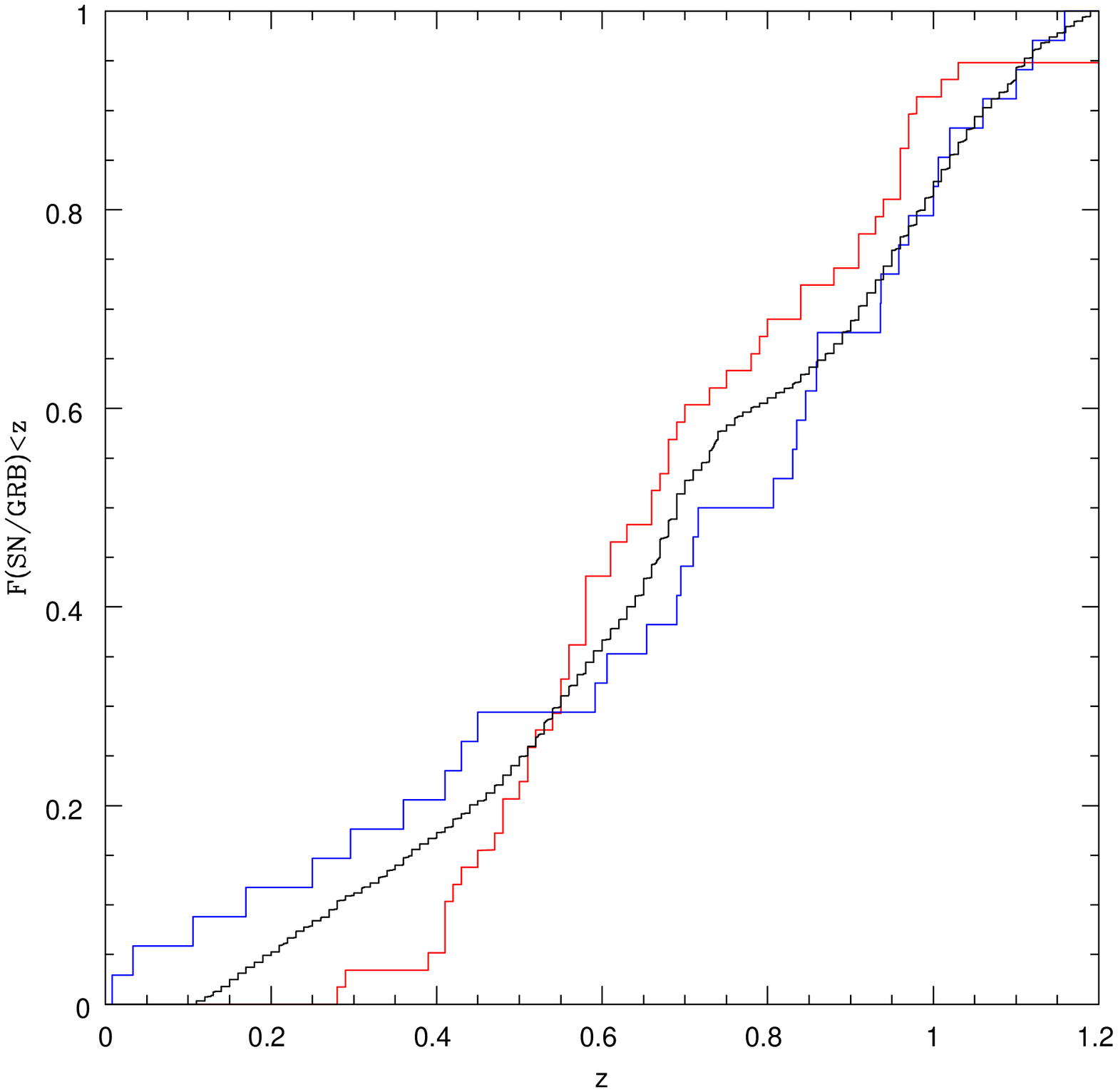}
  \caption{The redshift cumulative distributions of the GRB (blue) and SN (red) samples used in this paper. To provide similar redshift
    distributions we only consider GRBs with $z<1.2$. The redshift distribution of ~6900 MUSIC field galaxies is plotted in black.}
  \label{fig:zdist}
\end{figure}

\subsection{GRB host galaxies}
The mean redshift of GRBs in the {\em Swift} era is $\rm \sim 2.5$ \citep{2006A&A...447..897J}, however 
a number of GRB host galaxies have been observed at redshifts across the same, or very similar range as that of
the GOODS CCSN sample. To approximately match the redshift distributions we use all GRB host
galaxies at $z < 1.2$. Images of the resulting sample, which have {\it HST} observations, are shown in
Figure ~\ref{grbmos}, the subset of the hosts for which we present {\it Spitzer} fluxes is shown in Figure ~\ref{IRmos}. A comparison of the resulting redshift distributions is shown in Figure ~\ref{fig:zdist}.
Using this sample enables us to create a consistent dataset for CCSN and GRB hosts to perform the analysis on. This 
is crucial for us to be able to compare the results in a methodical way. 
The majority of the photometry for GRB host galaxies fitted here is taken from F06 
and \cite{2008arXiv0803.2718S}. However, we have supplemented this data with 
{\em Hubble Space Telescope} observations of
4 GRB host galaxies at $z<1.2$ (GRB/XRF 050416, GRB 050525, GRB 060218 and GRB 080319B\footnote{Host photometry extracted after subtraction of point source, see also \cite{2008arXiv0812.1217T}}) and {\em Spitzer} IRAC observations
of a further 13 hosts. 
The use of {\it HST} allows us to resolve
these galaxies and thus compare not only their luminosities but physical sizes. {\it HST} data were reduced
in the standard fashion via {\tt multidrizzle}, and magnitudes and radii were determined following
the method described in F06. See section \ref{phot:irac} for a description of the
IRAC photometry.
 Although deep imaging across multiple bands is
available we do not include the ambiguous GRBs 060505 and 060614, whose membership of the
long duration category of GRBs is controversial 
\citep[e.g. see][for a discussion of different viewpoints]{2006Natur.444.1044G,2006Natur.444.1047F,2008ApJ...676.1151T,2008ApJ...677L..85M}

Although the above selection allows us to largely remove any redshift bias from the observed
population, there do remain important selection differences between the GRB and CCSN host
population. Whilst these are difficult to quantify they should be considered before
conclusions regarding the two populations are drawn. The first effect is that the CCSN have
been located in a blind field search, and have a wide range of complementary data. This means
that it is possible to derive at least a photometric redshift for every CCSN within the sample. 
In contrast there are a number of very faint GRB host galaxies, which do not have spectroscopic 
redshifts, and have insufficient bands for photometric redshifts to be plausible. Should these
lie in the range of redshift we consider here ($z<1.2$) their non-inclusion would
tend to bias the observed population to higher luminosity. Indeed, even for the systems with
measured redshifts, the majority of our low-$z$ sample, $\sim 28$ from 34 come via emission line measures
in their host systems, rather than absorption lines in the afterglow, which may well create
a bias towards brighter hosts, and will be considered in more detail later.
In a similar spirit we have included GRBs with hosts identified both by their optical afterglows and
where the X-ray afterglow is sufficient to unambiguously locate the host, however it should be noted
that bursts with particularly faint optical afterglows (by dust extinction) could be missed from the sample.

Finally, there are a number of host galaxies at known redshift (GRBs 980326, 990705, 991216, 050416A, 050525A, 050824 and 051016B),
which have observations in a single photometric band, precluding a detailed analysis of their spectral energy
distributions. Excluding these would create a further bias within our samples, and so, rather than
omitting them we derive physical parameters by assuming they can be fit with the spectral 
template which provides the best bit to the majority of the GRB hosts. Although
this produces potential systematic errors into our analysis (for example the fainter galaxies
may typically have different colours than the brighter systems where our templates are derived) it
is preferable to their complete omission.

\subsection{GOODS-MUSIC: A comparison sample}

The GOODS-MUSIC (MUltiwavelength Southern Infrared Catalog) \citep{2006A&A...449..951G} includes 
photometry ranging from U-band (2.2ESO and VLT-VIMOS) to the 8 $\mu m$ IRAC band. Of the $\sim 14 000$
 objects listed in the catalog, we select $\sim 6900$ non-stellar, non-AGN objects with $0.1<z< 1.2$ 
(redshift either spectroscopic or photometric) as a field galaxy comparison sample to the GRB and CCSN populations. 
The object selection for the MUSIC catalog is made in the ACS z-band with a secondary selection made in the 
Ks-band to obtain a higher completeness. The limiting magnitudes are reported to be $z_{lim} \sim 26$ or $K_{lim} \sim 24$ (AB magnitudes)
at a completeness level of 90 \%.

Although this is a magnitude limited catalog, whereas the GRBs and CCSNe are are detected independent of
host magnitude, we consider this a good sample of field galaxies at similar redshifts to those
of the GRBs and CCSNe described above. It should also be noted that method of selecting
the MUSIC galaxies does not bias towards highly starforming galaxies like the selection based
on core-collapse events does. The MUSIC galaxies are hence bound to give a representation
of all Hubble types, i.e. include starforming spiral and irregular galaxies as well as passive
elliptical galaxies.   
 
% ############################################################################################################
%
% ############################################################################################################

\section{Photometry}

Image data from GOODS is used to acquire photometry in up to 12 bands. B, V, I and Z bands are
taken from \textit{Hubble}'s Advanced Camera for Surveys (ACS). Near infrared J, H and K bands from ground based 
Very Large Telescope (VLT) using the Infrared Spectrometer And Array Camera (ISAAC). 
Infrared images come from \textit{Spitzer}'s InfraRed Array Camera (IRAC) 
at 3.6, 4.5, 5.8 and 8 $\mu m$ wavelength. Further infrared magnitudes at 24 $\mu m$  ({\it Spitzer} MIPS) are 
adopted from \cite{2005ApJ...635.1022C}.
The ACS data comes in  high resolution (0.03 arcseconds per pixel) drizzled images. We use the  online cutout-service 
\footnote{http://archive.stsci.edu/eidol.php} to extract only the galaxy and it immediate surroundings from 
the larger mosaic image. 
The \textit{Spitzer} images are lower resolution and one image of manageable size covers
 the entire field. 

Photometry on the ACS images for the 16 hosts in the original sample (F06) is initially done 
with the qphot package in {\sc iraf}. We then compared this photometry with the GOODS source
catalog \citep{2004ApJ...600L..93G}, and finding a good agreement between them, we adopted catalog values
for all of the hosts.
Photometry on the ISAAC data, J,H and K bands was also checked for consistency between automatic source detection
via {\sc SExtractor} \citep{1996A&AS..117..393B}
and manual aperture photometry, 
after which we create our own source catalog, and adopt values from this for all of the hosts.
Due to the high amount of blending in the IRAC bands, automatic source detection is more challenging 
than for the optical and NIR bands. Photometry of the IRAC data is performed by hand, see below for a more detailed 
description. 

In addition to photometric data we also extract measured radii from the GOODS catalogue values. 
These are converted into physical sizes using our assumed cosmology ($\Lambda CDM$, $\Omega_M$ =0.27, $\Omega_{\Lambda} = 0.73$, $H_0 = 71$ km s$^{-1}$ Mpc$^{-1}$). 

The majority of the host galaxy photometry for the GRB host galaxies is collected from the GHostS project, where the photometry is
compiled from numerous sources, see \cite{2008arXiv0803.2718S} and references within. 
  All photometry has been corrected for galactic extinction following \cite{1998ApJ...500..525S}. 
  
\subsection{IRAC photometry}
\label{phot:irac}
The GOODS fields have been imaged in the {\em Spitzer} IRAC bands, from which we
have measured and report photometry for 56 of the CCSN hosts in Table \ref{tab:phot2} of the Appendix.
A number of GRB hosts have also been imaged in the IRAC bands, in addition 
to the reported magnitudes collected from the GHostS project. We have analysed these images and report
26 new $3.6\mu m - 8.0\mu m$ magnitudes or  magnitude limits for GRB hosts in Table \ref{tab:grb_irac} of the Appendix.

Note that, due to the amount of blending between sources at IRACs resolution, for some galaxies reliable photometry could not be achieved. 
In these cases the catalog entry is left blank. 

The GOODS observations have been mosaiced and drizzled to a pixel scale of 0.6 arcsec/pixel, limiting magnitudes are ~24-25 depending
on the IRAC band and extent of the source, as estimated from {\it HST} imaging.
The GRB observations are reduced by the standard IRAC pipeline, and have the native pixel scale of 1.2 arcsec/pixel. Limiting magnitudes
are ~ 19-23 depending on exposure times and bands of the individual observations. 
 
The photometry is performed using the python package PyFITS provided by STScI, to extract (normal extraction) the flux inside a 
circular aperture with sub-pixel accuracy.
The background is measured from blank apertures outside the host, which also provide the background standard deviation 
for determination of limiting magnitudes. Quoted limits are 3-sigma.

At the resolution of IRAC, the majority of the hosts are unresolved; 
in which case we use small aperture photometry  and aperture corrections according to the
official IRAC calibration (for the GRB hosts) or as determined from the curve of growth (CCSNe in the GOODS mosaic).
 If the source emission is determined to have a FWHM larger than the FWHM of the PSF, we extract the photometry from a large aperture
enclosing all of the flux.

% ############################################################################################################
%
% ############################################################################################################

\section{Spectral Energy Distribution fitting}

The collected photometry covering wavelengths from 0.4 $\mu m$~(ACS B-band) to 24 $\mu m$ 
(\textit{Spitzer} MIPS), allows us to fit template spectral energy distributions that are close 
representations of  the true SED within these limits. 
Redshifts for the CCSN hosts are determined spectroscopically in 41 cases and photometrically in 17.
Spectroscopic redshifts are adopted either from \cite{2004ApJ...613..200S} where available, or by
querying the Team Keck Treasury Redshift Survey (TKRS) \citep{2004AJ....127.3121W} for the GOODS north field, 
or the GOODS/FORS2 release 3 \citep{2005A&A...434...53V,2006A&A...454..423V,2008A&A...478...83V}  online redshift catalog in the south field.
Photometric redshifts are calculated with the {\sc HyperZ} photometric redshift code 
\citep{2000A&A...363..476B} \footnote{Consistency is checked using objects overlapping with the MUSIC catalog.}. 
Our own SED fitting includes only two degrees of freedom: a wavelength independent flux proportionality, and and 
a reddening inside the host galaxy that is wavelength dependent and 
calculated in the host restframe. The reddening curve is adopted from \cite{2000ApJ...533..682C} 
which
is derived to suit actively starforming galaxies.

Template spectral energy distributions are collected from the literature. They include both observed
SEDs of local galaxies and SEDs produced with various spectral synthesis codes.
Mean templates for local ellipticals and spirals galaxies are adopted from \cite{1980ApJS...43..393C}. 
Synthetic GISSEL98 spectra ranging along the entire Hubble sequence are adopted from 
\cite{1993ApJ...405..538B}, and synthetic fits for local galaxies ARP220, HR10, M51, M82, M100, 
NGC 6090 and NGC 6946 are adopted GRASIL spectral libraries of \cite{1998ApJ...509..103S}.
We also include GRASIL synthetic templates fitted for submm selected GRB hosts by \cite{2008ApJ...672..817M}.

\begin{figure*}
  \includegraphics[scale=0.65]{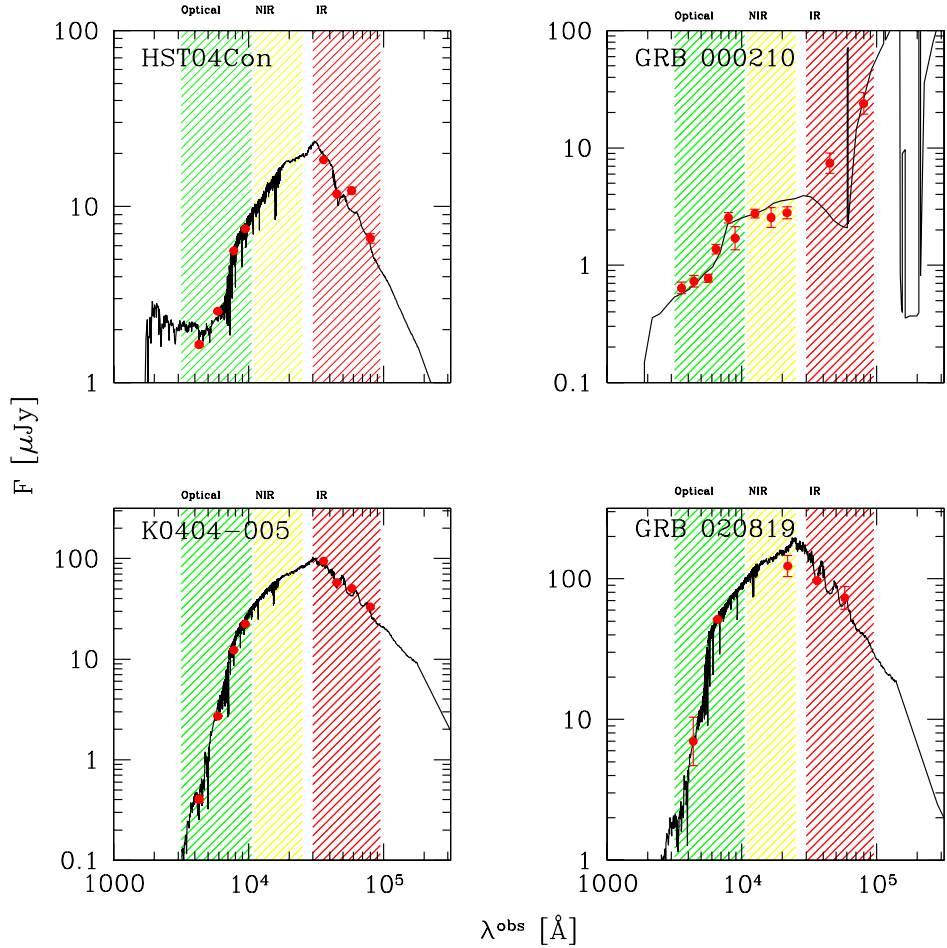}
  \caption{Example spectral energy distribution fits. Wavelengths are in  the observed frame. Host galaxies of SNe HST04Con and K0404-005 have absolute
    V magnitudes of -21.37  and -22.53  respectively. The hosts of GRBs 000210 and 020819 have
    absolute magnitudes of -20.07 and -21.93 respectively}
  \label{fig:SED}
\end{figure*}

\begin{figure}
  \includegraphics[scale=0.43]{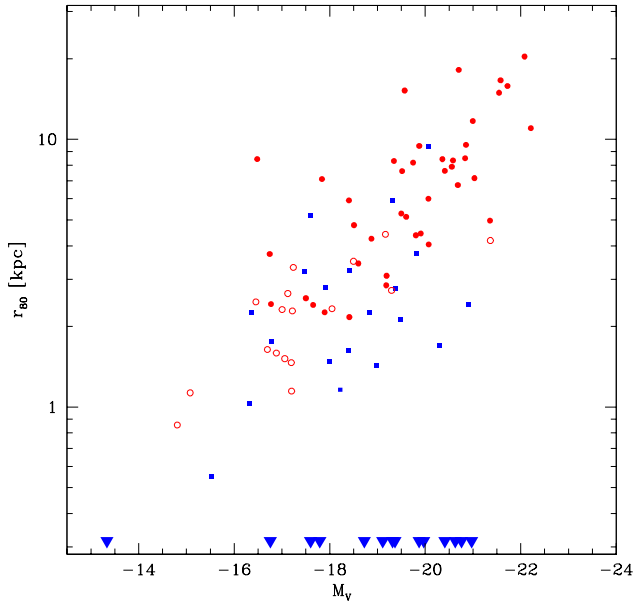}
  \caption{80 \% light radius versus absolute V band magnitude for GRB hosts (blue squares), CCSN hosts (red points, filled
    for hosts with spectroscopic redshifts).
    Blue triangles on the bottom axis are the absolute magnitudes for GRB without a measured radius (i.e. those without {\em HST} imaging).}
  \label{fig:Msize}
\end{figure}

The best fit is given by minimising 

\begin{equation}
\chi^2 = \sum_{i=1}^{N_{filter}} \left ( \frac{f_{i,obs}-b \times f_{i,template} \times 10^{\frac{k(\lambda)A_v}{R_v}}}{\sigma_{i,f} } \right ) ^2
\end{equation}

with respect to the scaling parameter b, and the reddening parameter $A_v$. The reddening curve $k(\lambda)$ and $R_v=4.05$ are fixed 
by the reddening law. The optimum SED template is  
transformed to its restframe and analysed to estimate physical 
parameters of the host galaxy.
For wavelengths between two photometric bands this means an interpolation that is more secure than a
linear interpolation or assuming a globally flat SED. Some examples of our SED fits are shown in 
Figure ~\ref{fig:SED}. Having determined the best fitting spectral templates we derive absolute 
magnitudes in given photometric bands by integrating the spectrum over the response function
of the filter. In Figure ~\ref{fig:Msize} we plot the derived $M_V$ values against the
radii of each host galaxy.

Below we describe in brief the parameter-SED relations
we use to estimate stellar mass content ($M_{\star}$) of the hosts, their star formation rates (SFR) and 
metallicities ($12+\log{O/H}$). Note that the spectral energy distributions are corrected for internal extinction added
in the fitting procedure when estimating these properties.

% ############################################################################################################
%
% ############################################################################################################

\section{Deriving Physical Parameters}

\subsection{Stellar masses}

The stellar component of the total  mass in a galaxy, $M_{\star}$, can be estimated using the rest frame 
K-band luminosity, which
samples the old stellar population with a much weaker contribution from hot and massive short lived stars.
We note that some caution has been suggested when using this method on stellar populations 
dominated by young to intermediate aged stars, as red supergiants can become a significant
source of enhanced K-band luminosity, and thereby lead to an overestimate of the stellar mass, e.g. \cite{1995ApJS...96....9L}.
A standard method of mass estimation is the mass to light ratio, where one assumes a proportional 
relationship between the stellar mass and the K-band luminosity. \cite{2006ApJ...653L..85C} 
prescribe $M_{\star}/L_K \sim 0.1$ for the GRB host galaxies in their sample. Our SED fits to these 
same galaxies give stellar masses in good agreement with the results of \cite{2006ApJ...653L..85C}. 
Here we have chosen to estimate the stellar masses with the relation of \cite{2009ApJ...691..182S},

\begin{equation}
\log M_* = -0.467 \times M_K -0.179
\end{equation}

which is calibrated on the basis of GRB hosts. (See also \cite{2004Natur.430..181G} for details on this mass calibration.)

\subsection{Star formation rates}

While the K-band luminosity is an indicator of the old stellar population in a galaxy, 
the U-band luminosity samples the SED contribution from the hot, massive and hence newly formed 
stars. Following \cite{1998ApJ...507..155C} we estimate the SFR by,
  
\begin{equation}
  SFR_U(\mathrm{all})=\frac{8.8 \times L_U}{1.5 \times 10^{22}Whz^{-1}} \mathrm{ M_{\odot}  yr^{-1}}.
\end{equation}

Where we introduced a factor $8.8$ to correct from $SFR_U(M/M_{\odot}>5)$ to account for all star formation.
It should be noted that this SFR is not model independent, but rather it assumes a certain 
initial mass function (IMF). \cite{1998ApJ...507..155C} assume a Salpeter IMF.
Both stellar masses and star formation rates may be inaccurately estimated
if the IMF is strongly deviating from that of Salpeter. Though is more likely to agree well with 
CCSN-like hosts that commonly are spiral galaxies, a low mass, metal poor galaxy, initially 
expected to be a GRB host, could have a more pronounced top heavy IMF.

\begin{figure}
  \includegraphics[scale=0.4]{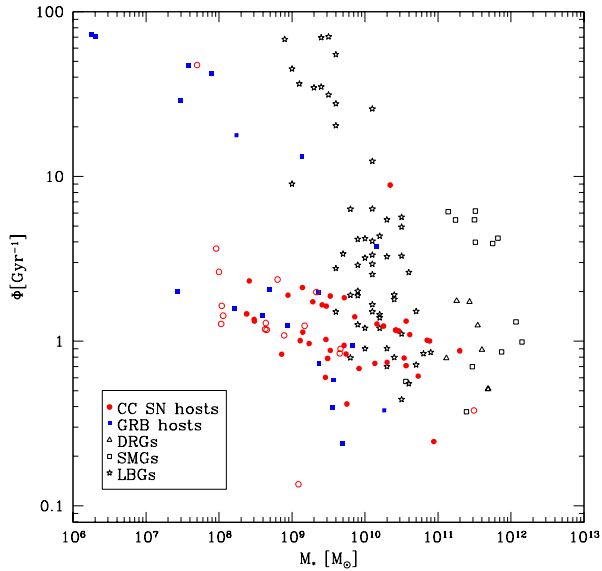}
   
  \caption{Specific star formation rates versus stellar mass for GRB hosts (blue squares), CCSN hosts (red circles, filled
    for hosts with spectroscopic redshifts)
    and  a selection of distant red galaxies (DRGs), Sub-mm
    galaxies (SMGs) and Lyman-break galaxies (LBGs) compiled by Castro Cer\'on et al. (2006)}
  \label{fig:SSFR}
\end{figure}

Further useful quantities are the specific star formation rate $\Phi$, 
\begin{equation}
\Phi=\frac{SFR}{M_{\star}},
\end{equation}
and the star formation surface density $\Sigma$,
\begin{equation}
\Sigma=\frac{SFR}{\pi r_{80}^2},
\end{equation}
-- star formation per unit stellar mass and unit area in the galaxy respectively.  Since these indicate how intense the
star formation is, they are in some regards a more interesting parameters to study than the
SFR itself. GRB hosts are believed to have high SSFR in general, as the presence of GRB itself
is evidence of the formation of massive stars. Indeed this is supported by \cite{2006ApJ...653L..85C} who
place the SSFRs of four $z \sim 1$ GRB hosts amongst the highest observed. 
In Figure ~\ref{fig:SSFR} we plot the SSFRs vs the masses for the GRB and CCSN hosting galaxy populations, as 
well  as a selection of other high-z galaxy populations.
In addition to the SSFR, we also define the surface SFR, $\Sigma$, as the SFR per unit area of the galaxy. 

\subsection{Metallicities}

The role of progenitor metallicity in determining the outcome of
massive-star core collapse has been discussed by various authors. With the difficulties in making direct measurements of the metallicity at high redshift, mass or luminosity are commonly used as proxies. The existence of a  relationship between galactic stellar mass and its metallicity has been known since \cite{1979A&A....80..155L} published their results
based on a sample of eight local galaxies. Their conclusion that low stellar
mass galaxies also have lower metallicities, has since been confirmed
and extended by using the much larger samples of local galaxies allowed by the
SDSS, e.g. \cite{2004ApJ...613..898T}. 
The origin of the M-Z relation is still under investigation. 
Loss of metal enriched gas via galactic winds, accretion of low metallicity
gas from the IGM, or lower starformation efficiencies in low mass galaxies could all effect the metallicity, and have been suggested as 
possible explanations, see e.g. \cite{1974MNRAS.169..229L} and \cite{1995ApJ...454...69P}.

\cite{2005ApJ...635..260S} calibrate the following mass-metallicity (M-Z) relationship using 69 
Gemini Deep Survey and Canada-France Redshift Survey galaxies with redshifts between 0.4 and 1,

\begin{equation}
  12+\log{\rm (O/H)}=0.478 \log{M_{\star}}+4.062
\end{equation}

This M-Z relation is claimed to be an improvement from the use of luminosity-metallicity
relations ($\sim 0.2$ dex scatter), largely due to the small variations through the galaxies 
evolution in the K-band luminosity used to estimate the stellar mass in the galaxies. While 
short starburst and star formation history modify the B- and V-band luminosity greatly, the 
K-band remains relatively constant.

% ############################################################################################################
%
% ############################################################################################################

\section{Locations}
In addition to their galactic environments the local scale environments of GRBs and SN can
also provide strong constraints on progenitors. If spatially resolved spectroscopy is
available then the chemical evolution of the progenitor region can be probed directly, 
however, this is only possible in a handful of cases 
\citep[e.g.][]{2008A&A...490...45C}. In the absence of detailed spectroscopy the luminosities
of the region containing the transient can also be diagnostic \citep[e.g.][]{2008MNRAS.387.1227O}. 
These luminosities
can be investigated both in relation to the overall host galaxy, and in absolute terms. 
Fruchter et al. (2006) developed a pixel statistic, where the galaxy is defined by
adjoining pixels above some signal to noise limit. These pixels are then sorted
into ascending order, and the pixel containing the GRB or SN is located in this ranked list. 
It is then possible to record a simple statistic -- the fraction of host light in pixels of
equal or lower surface brightness than the pixel containing the GRB or SN. 
This technique has the significant advantage that it provides information on the location
of a given transient which is broadly independent of the morphology of the galaxy. This
is particularly important for high redshift hosts, which often show disturbed and irregular
morphologies. 
The analysis of Fruchter et al. (2006) showed that GRBs are significantly more
concentrated on their host light than the SN, and this is naturally interpreted as GRBs
originating from more massive stellar progenitors \citep{2007MNRAS.376.1285L}. A similar
result was obtained by \cite{2008ApJ...687.1201K} for type Ic supernovae, also suggesting
a higher mass origin for these systems \citep{2008ApJ...689..358R}. 

We have extended the analysis of Fruchter et al. (2006) to include more recent CCSN and
GRBs. The GRB sample is only moderately enhanced from the sample of Fruchter et al (2006),
since the number of bursts with accurate positions and {\it HST} observations is not dramatically
larger in the {\em Swift} era. However, the CCSN sample has increased by a factor of 4.
To derive locations for the transients we co-align images taken at different epochs, one
in which the SN/GRB is bright, and the other where it absent (for GRBs this is normally a very late
time image, while for SN it is frequently a pre-explosion image). We then perform a direct
subtraction of the two {\it HST} images and centroid on the variable source. We then create
a galaxy mask via {\sc SExtractor} and locate the pixel containing the GRB/SN in its cumulative distribution. 

An alternative approach 
is to investigate the surface brightness of these pixels, and thus of the region of the host galaxy 
containing the GRB or SN. By doing this, one can make a direct comparison of the local luminosities
of GRB and CCSN, essentially measuring the luminosity of the populations which host them. 
Since the luminosity of a given star is roughly proportional to the cube of its mass $L_B \propto m_{star}^3$, the mass (and hence age) of the stellar population dominates this statistic, more strongly
than, for example, stellar number counts, where $L_B \propto N_{stars}$. Since the GRB and CCSN host galaxies lie at similar 
redshifts the physical scales probed by this are comparable\footnote{A pixel is roughly 150-200pc 
on a side}. 

We perform this analysis using the full sample of 58 CCSN shown in Table ~\ref{tab:CC}. For
the GRBs, we utilise a  subset  of the sample as F06, where the
burst lies at $z<1.2$ with a positional accuracy of $\lsim 0.08$\arcsec, such
that the location of the burst was known to better than the {\it HST} (WFPC2 or ACS) PSF, and thus
the images did not require additional smoothing to emulate the observation of the
host at the resolution of the error region.  We have calculated 
the true surface brightness of the pixel that contained
the CCSN or GRB event in units of $L_{\odot}$ kpc$^{-2}$ for a subsample of hosts. To 
account for the differing redshifts of our sample we make K-corrections to these
values assuming that the locations of the transient have the same colours indicated by 
global photometry of the host galaxy. This introduces a degree of error since the colour
mapping across the galaxy is unlikely to be constant. However, the signal to noise of
individual pixels is normally too low to place strong constraints on the pixel colours. We
note that the application (or not) of this correction does not significantly impact our results. 
Our resulting distribution in shown in Figure ~\ref{fig:cdf2}, and confirms that not only do GRBs trace a high power
of light within their host galaxies, but also that GRB hosting regions are much brighter than those
which host a CCSN. 

% ############################################################################################################
%
% ############################################################################################################

\section{Results}

The results of our analysis for CCSN and GRB hosts are shown in Tables ~\ref{tab:CC} and ~\ref{tab:GRB}, where 
we have tabulated the parameters derived from the fits 
(absolute magnitudes, star formation rates, stellar masses and metallicities) along with directly measured parameters
$(r_{80})$. The raw photometry used for the fits to the CCSN hosts is presented in Appendix 1. 
The median V band absolute magnitudes are -20 (CCSN) and -19.4 (GRB) respectively, median masses are 
$\rm 3 \times 10^9 \Mo$ (CCSN) and $\rm 1.3 \times 10^9 \Mo$ (GRB), median star formation rates and specific star formation 
rates are $\rm 3.6 \Mo yr^{-1}$ (CCSN), $\rm 1.6 \Mo yr^{-1}$ (GRB) and $\rm 1.2 Gyr^{-1}$ (CCSN), $\rm 1.2 Gyr^{-1}$ (GRB).

\begin{figure}
  \includegraphics[scale=0.35]{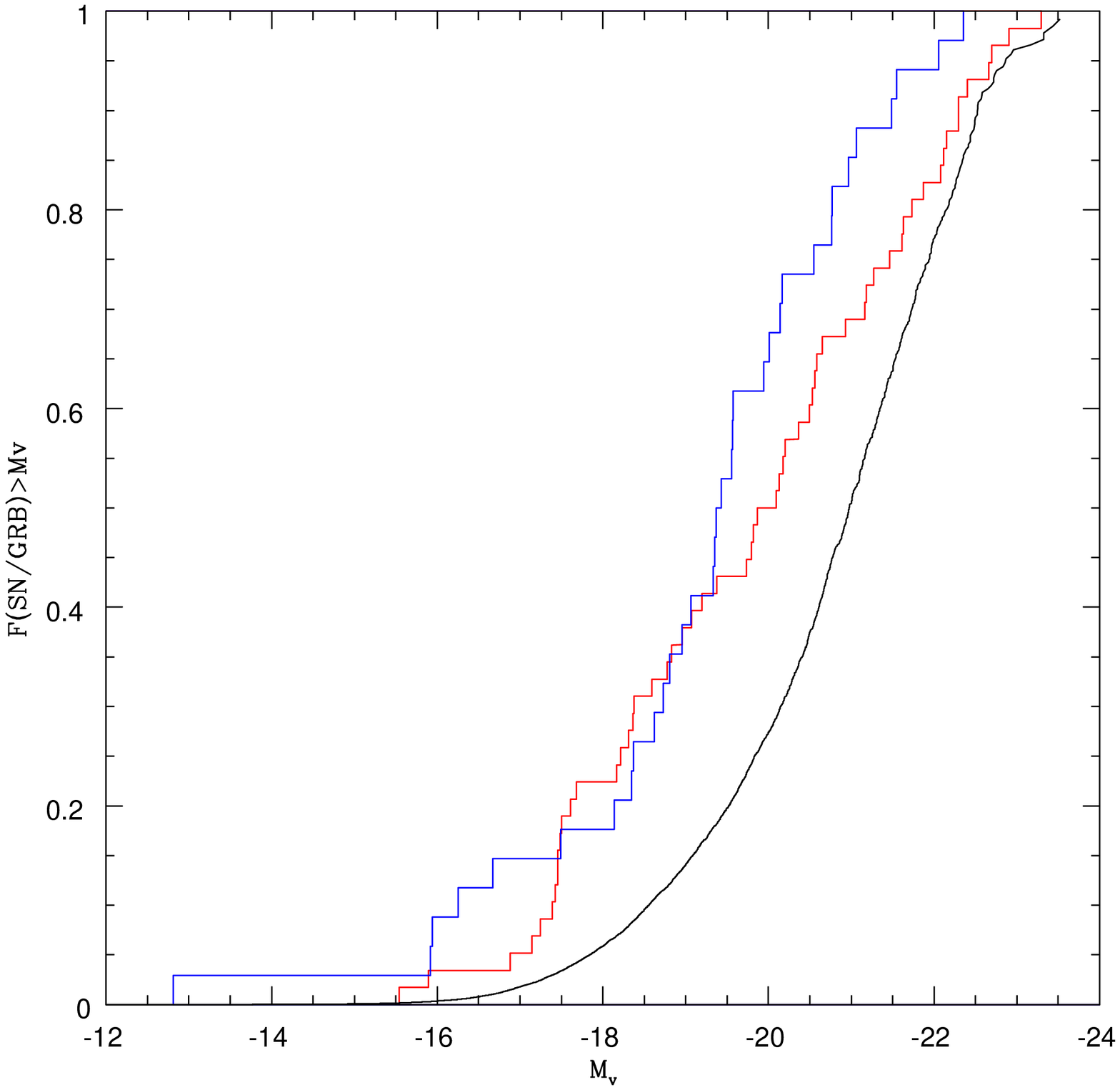}\\
  \includegraphics[scale=0.35]{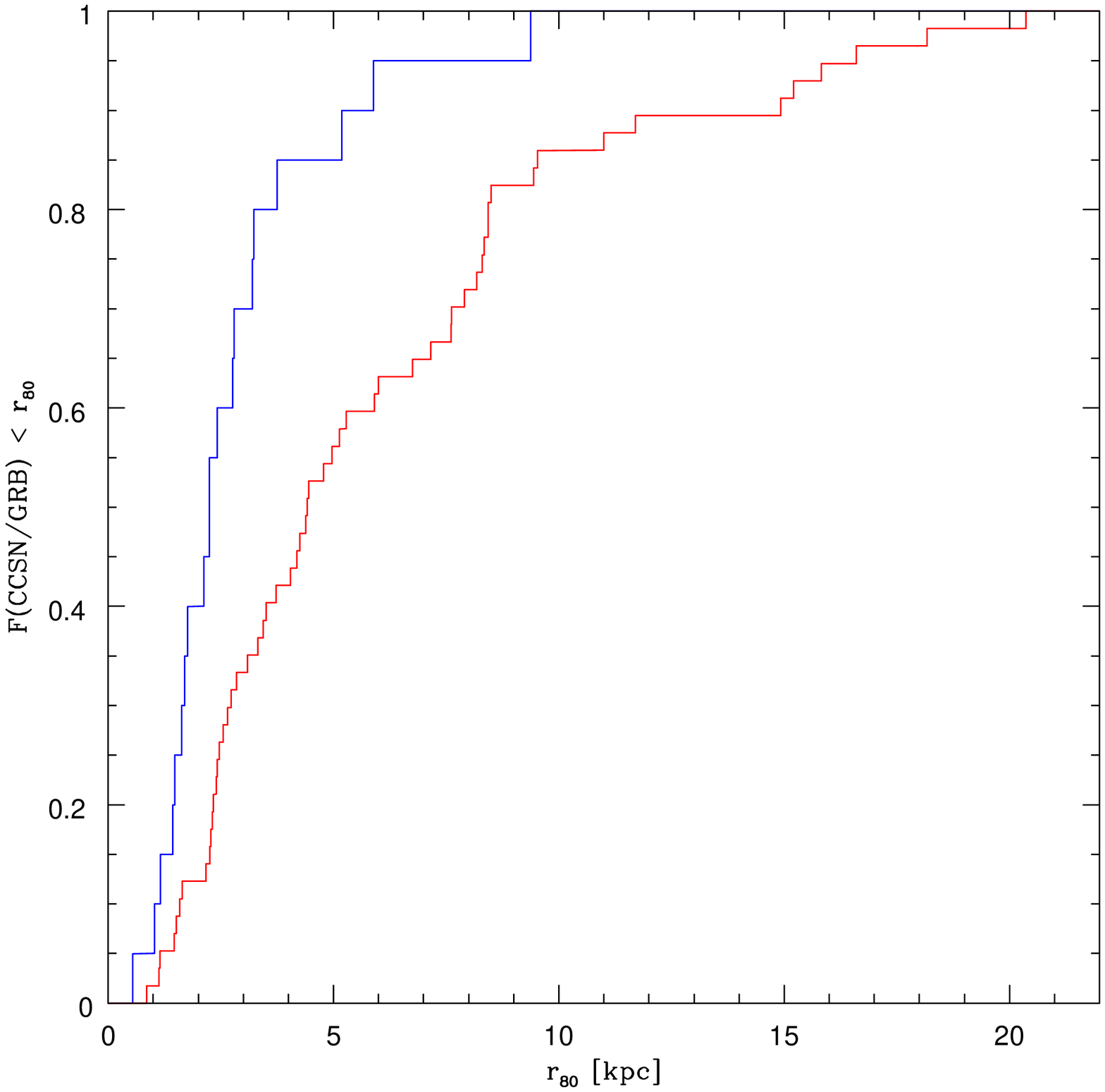}
\caption{
  Cumulative distributions of the absolute V-band magnitudes of GRB hosts (blue line), CCSN hosts (red) and
  the MUSIC field galaxy sample (black) with absolute magnitudes accumulated by luminosity.
  {\it \bf (Upper)} and 80\% light radius{\it \bf (Lower)}.
}
\label{fig:cdf0}
\end{figure}

\begin{figure*}
  \includegraphics[scale=0.35]{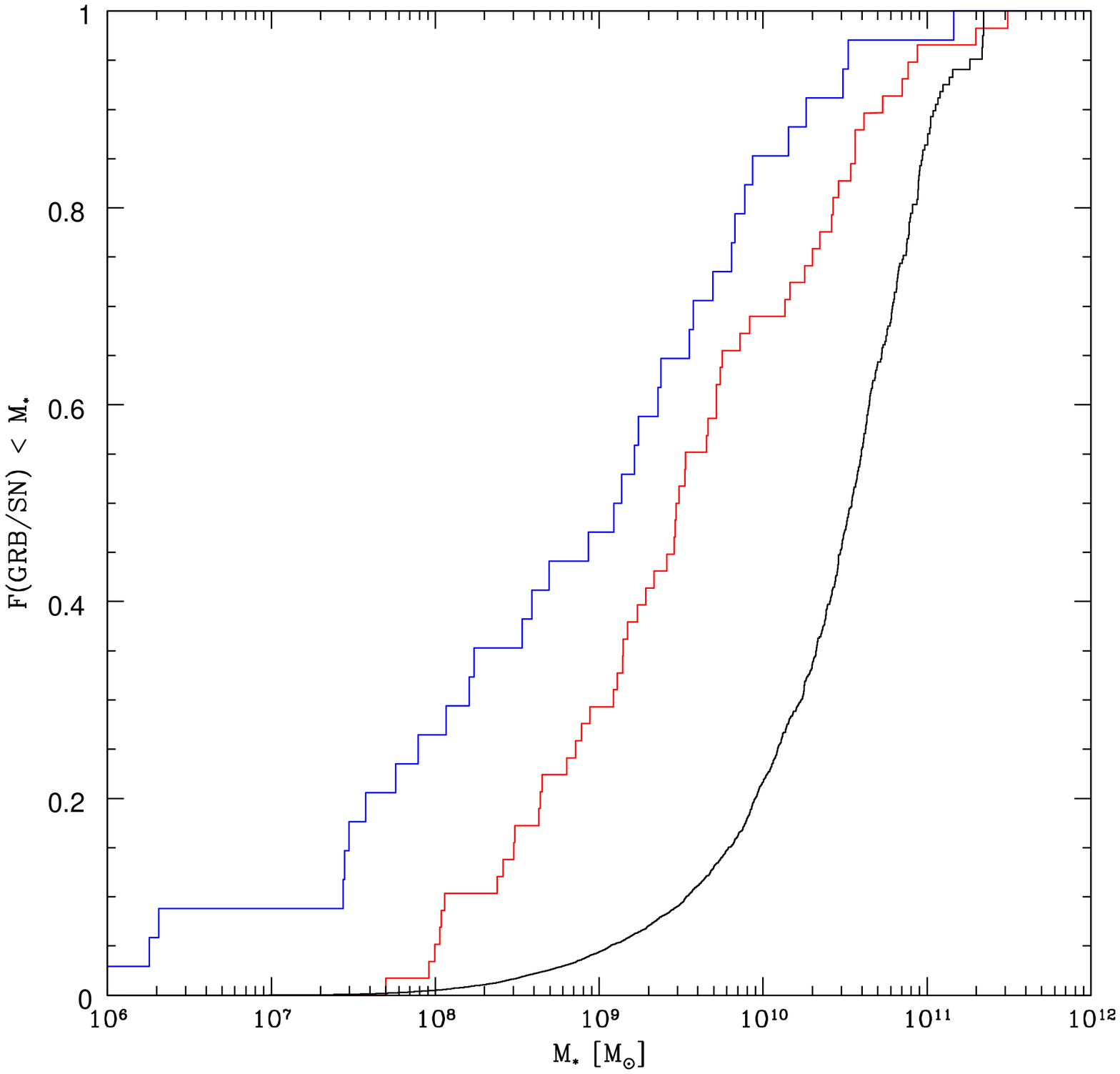} 
  \includegraphics[scale=0.35]{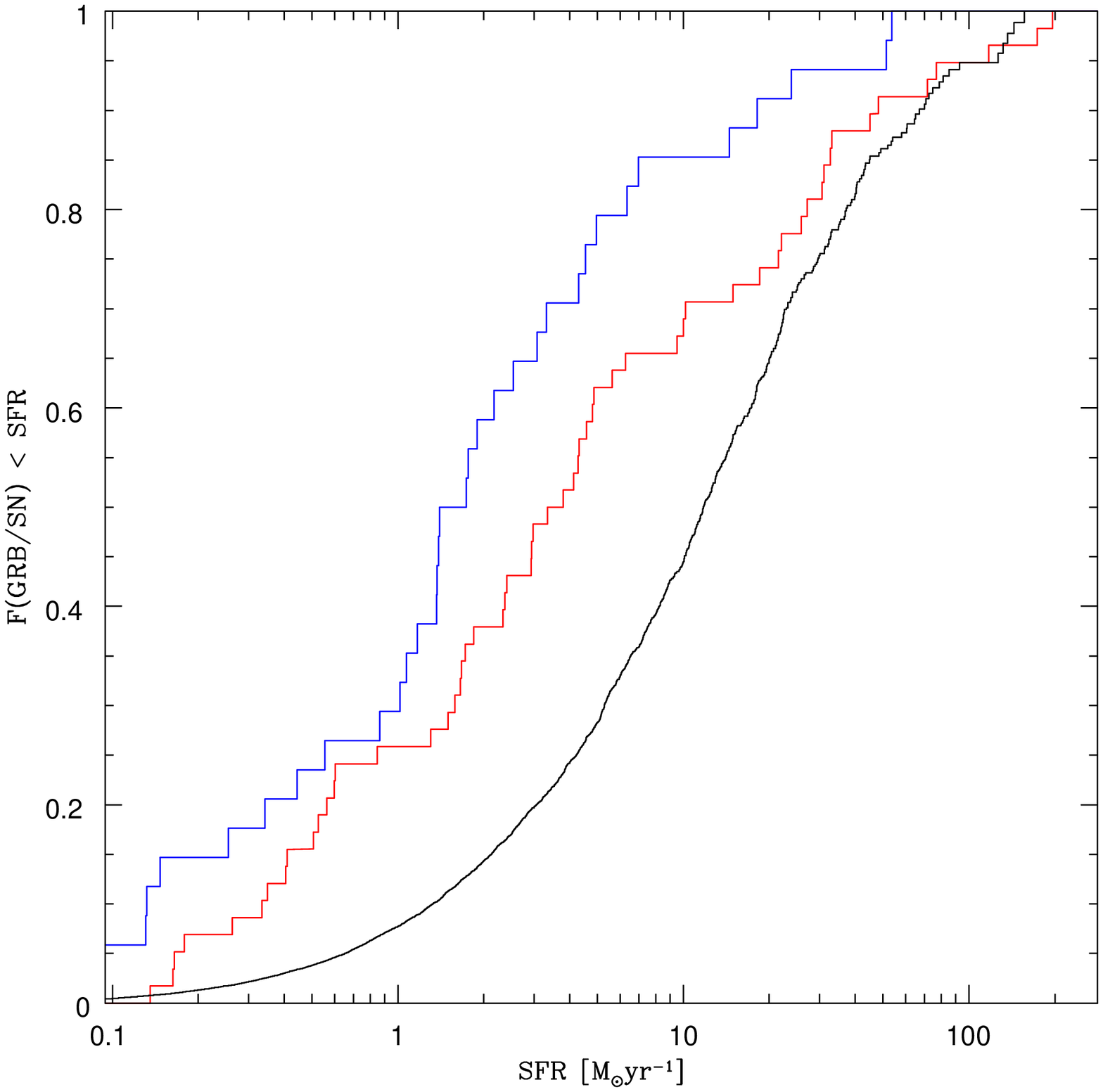}\\
  \includegraphics[scale=0.35]{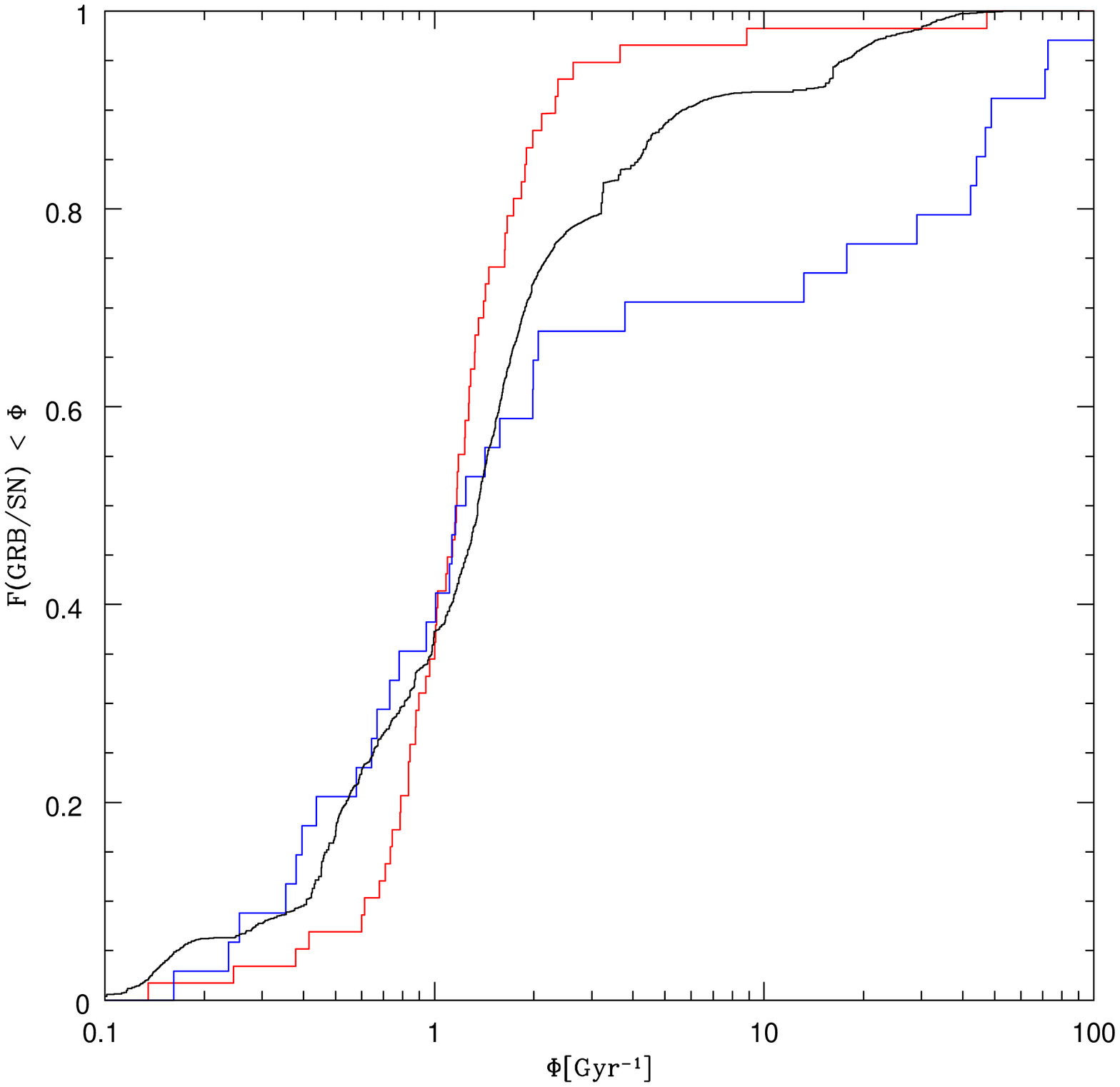}
  \includegraphics[scale=0.35]{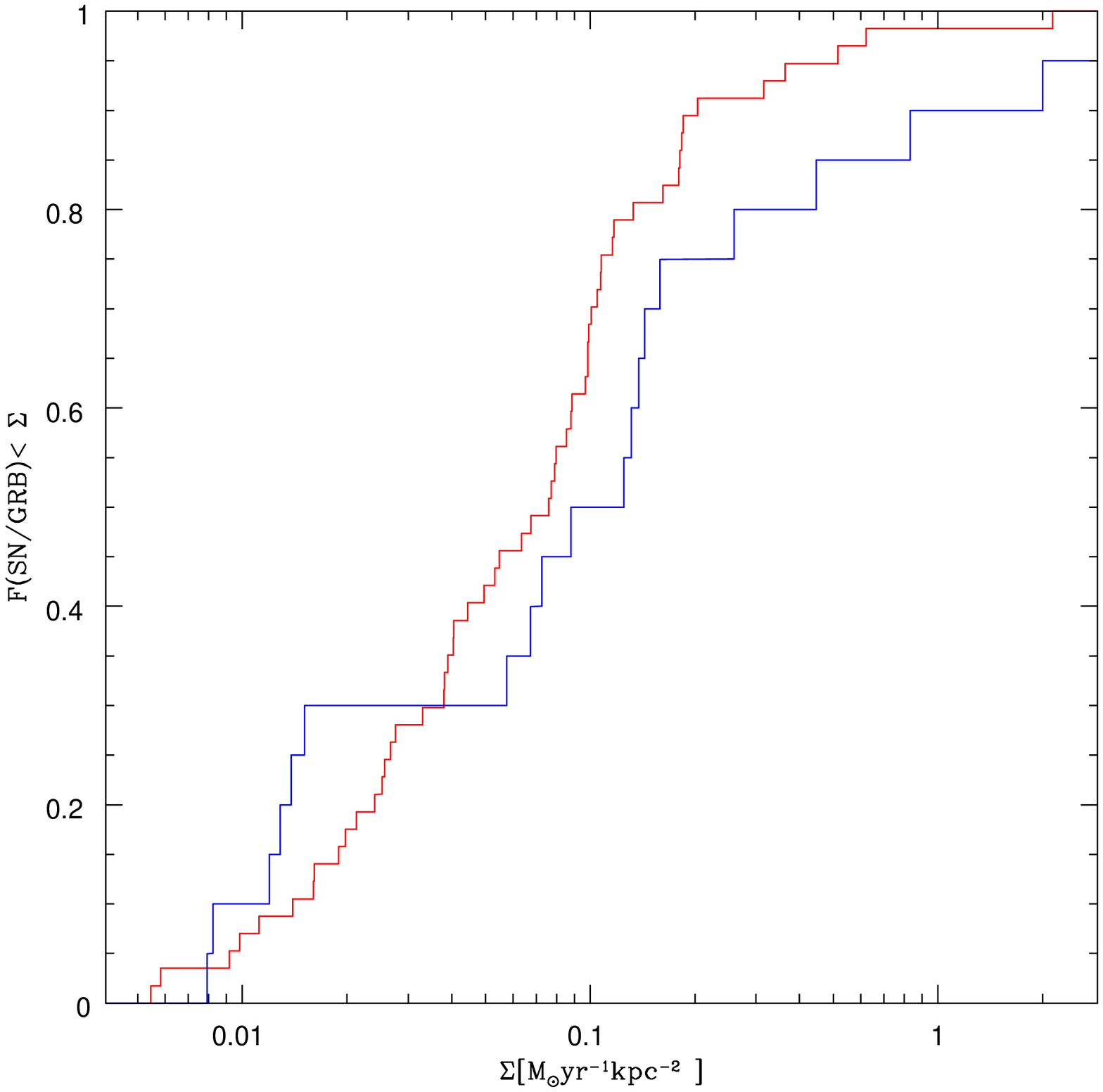}\\
  \caption{
    {\it \bf Top left:} 
    Cumulative distribution of CCSN (red) and GRB (blue) host 
    galaxy masses along with fractional mass distribution in field galaxies (black).
    Note that for CCSN and GRB we plot the fraction of number of galaxies, while for the field galaxies, we plot the fraction of mass. 
    {\it \bf Top right:} 
    Cumulative distribution of the star formation  rates. The field galaxy sample is weighted by the individual galaxies SFR.
    {\it \bf Lower left} 
    Cumulative distribution of CCSN and GRB specific star formation rates. The field galaxies SSFR is weighted by the SFR in each galaxy. 
    {\it \bf Lower right} 
    The surface star formation rates of GRB and supernova host galaxies, assuming a uniform distribution
    of starformation over $r_{80}$.
  }
  \label{fig:cdf1}
\end{figure*}

\begin{figure}
  \includegraphics[scale=0.35]{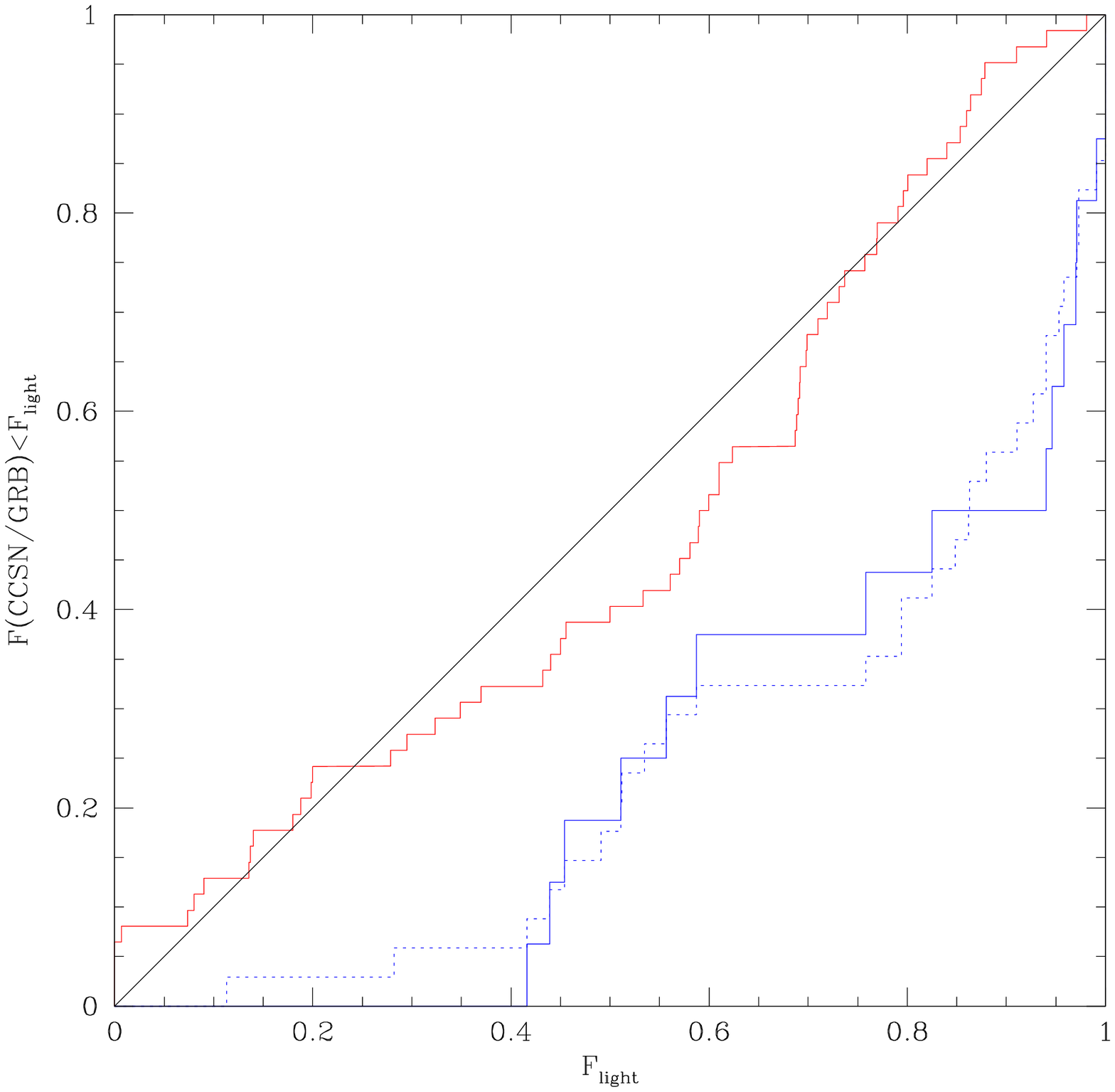}\\
  \includegraphics[scale=0.35]{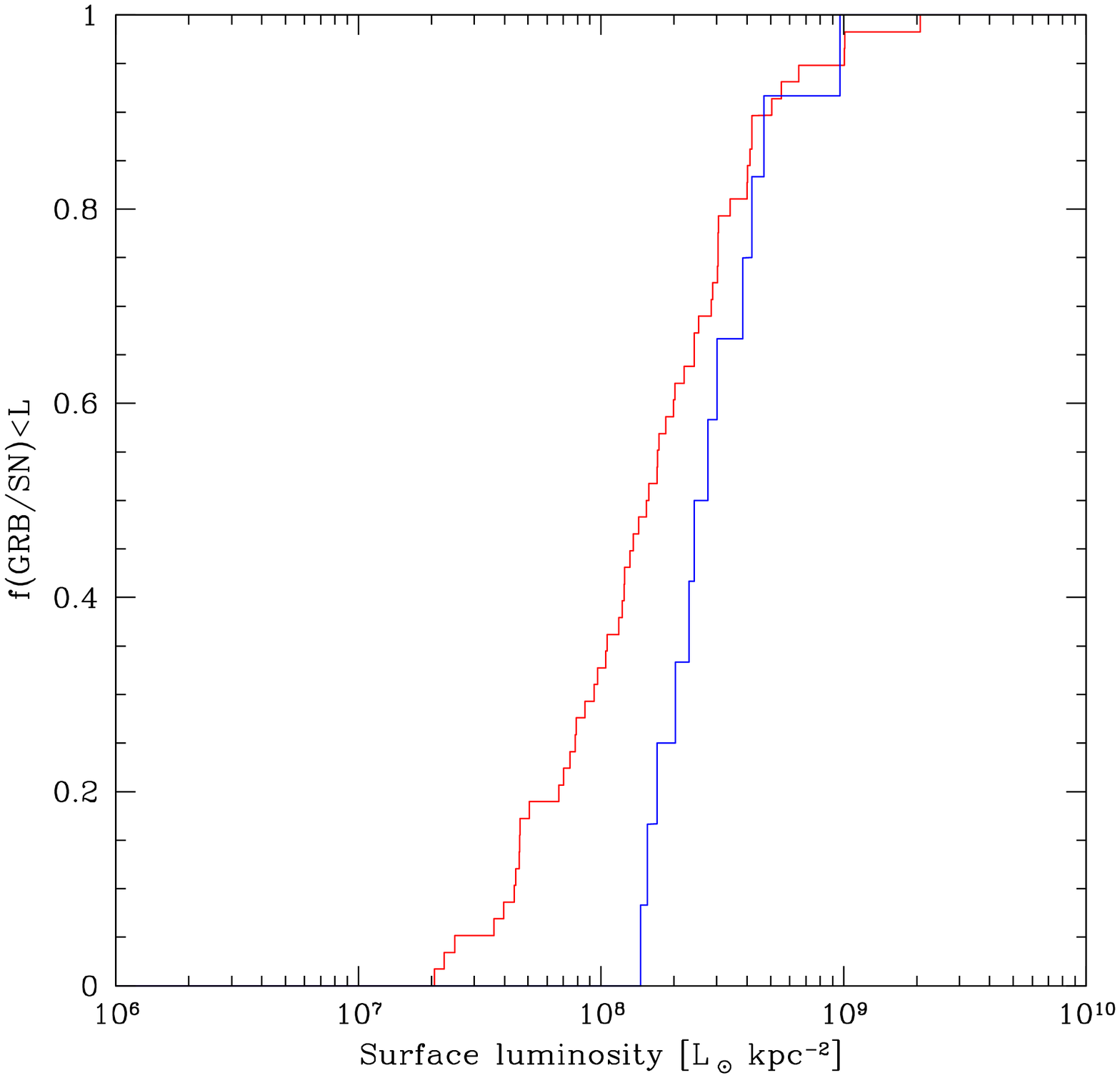}\\
  \caption{Local environmental properties of the GRB and CCSN sample. {\bf Upper:}, the locations of 
    SN (red) and GRBs (blue) on the light distributions of their host galaxy. The blue dashed line
    shows the locations for GRBs at $z<1.2$, while the solid line all bursts in the sample
    of Fruchter et al. (2006).  {\bf Lower:} The absolute surface
    brightness under the transient location in $L_{\odot}$  kpc$^{-2}$. Both in relative and absolute terms
    GRBs appear more concentrated on their host galaxy light.  }
  \label{fig:cdf2}
\end{figure}

We perform KS-tests on the cumulative distributions of all the parameters to formalise the probabilities 
that they are drawn from a single population. The KS probabilities are listed in Table ~\ref{tab:Pks}, and a selection 
of the cumulative distribution functions are plotted in Figures ~\ref{fig:cdf0} to ~\ref{fig:cdf2}.

We also compare the GRB/CCSN selected galaxies with the GOODS-MUSIC field galaxy sample. Since this sample is 
selected differently from the CCSN or GRB hosts, we cannot simply compare the field CDFs to the CCSN/GRB CDFs. 
Instead, for ${\rm M_{\star}}$  we accumulate the mass in every step so that the step height is proportional 
to the mass of each field galaxy instead of constant.
Hence, where the CDF for the CCSN/GRB hosts shows the number of galaxies with ${\rm mass<M_{\star}}$ the accumulated 
function shows the {\it fraction of total mass in the field} that is accounted for by galaxies with ${\rm  mass<M_{\star}}$.
The principle for the SFR and ${\rm \Phi}$ is the same, but${\rm \Phi}$ weighted by SFR instead of ${\rm\Phi}$ itself, i.e.
this distribution function shows what fraction of star formation occurs in galaxies less active than ${\rm \Phi}$. In plotting
the field galaxies in this way we would expect agreement between the field galaxy and GRB/SN curves if the
probability of a GRB occurring in a given galaxy were directly proportional to the SFR (or mass) of the galaxy.

Unlike previous work we do not find any statistically significant differences between the absolute magnitudes of the
GRB and CCSN host populations: the hypothesis that they are drawn from the same population is
accepted with probability $P_{KS} = 0.4$  for both $M_B$ and $M_V$, although the median $M_V$ of the CCSN hosts is a factor of 2 
brighter in luminosity than that of the GRB hosts. 
Also the restframe B-V colours of CCSN hosts are also similar to those of GRBs with a probability $P_{KS} = 0.2$. 

However, though the stellar masses and and star formation rates are also broadly comparable 
($P_{KS} = 0.12$ and $P_{KS} = 0.15$), when weighting the star formation by the galactic mass this suggest that
the specific star formation rates for GRB hosts are higher than for CCSN ($P_{KS} = 0.04$). 

A comparison of the radii of the two galaxy samples also suggests, at a high significance, that 
GRB hosts are smaller than those of CCSN ($P_{KS} = 0.003$). 
These results suggest that GRB hosts are on average smaller, and and more actively starforming than 
the CCSN counterparts. We also note the distribution of CCSN
surface luminosities ($\Sigma$), which essentially combines their size and luminosity, 
is higher than that of GRB hosts, although not at a statistically significant level, $P_{KS} = 0.14$

Further evidence for the difference between the progenitors of CCSN and GRBs comes from
their locations. Despite a relatively small sample of GRBs with highly accurate positions on their hosts it
is clear that they typically occur in regions of much higher surface brightness than CCSN, with the
median difference between GRB and CCSN hosting sites being a factor of 4 in surface brightness
($P_{KS} = 0.01$), and $P_{KS} = 5 \times 10^{-3}$ when comparing the relative brightness ($F_{light}$) of the explosion site.

\begin{table*}
  \caption{Name of the associated core-collapse  event, the redshift and quantities derived from the Spectral energy distribution fits; 
    Absolute magnitude in the V- and B-bands, star formation rate and stellar mass content. Hosts with only photometric redshift determination are 
    marked in italic. Note that $F_{light}$ and surface luminosity for bursts 2002fz to 2003N are calculated in the F606W filter, while the rest
  are in the F850LP filter.}
  \footnotesize{ 
    \centering
    \begin{tabular}{l| l l l l l l l l l l}
      \hline
      \hline
      SN name&  {\it z} &$\rm r^{80}$ &  $\rm M_V$&$\rm M_B$& $\rm SFR$ & $\rm \log  M_{\star} $& $\rm 12+$ & Surface Lum & $\rm F_{light}$\\
      &   &$\rm [kpc]$ &  AB mag& AB mag& $\rm [M_{\odot}/yr]$ & $\rm [M_{\odot}]$&$\rm \log(O/H)$ &${\rm [\log{(L_{\odot}kpc^{-2})}]}$& \\
      \hline   
\textit{2002fv} & 0.7 & 0.86 & -15.9 & -15.47 & 0.18 & 8.04 & 7.9 & 7.83 & 0.46 \\
2002fz & 0.84 & 11.7 & -22.08 & -21.64 & 45.01 & 10.61 & 9.14 & 8.2 & 0.59 \\
2002hs & 0.39 & 8.43 & -17.24 & -16.89 & 1.3 & 9.11 & 8.42 & 7.67 & 0.09 \\
2002hq & 0.67 & 16.6 & -22.66 & -22.22 & 76.78 & 10.88 & 9.26 & 8.16 & 0.37 \\
2002kb & 0.58 & 15.82 & -22.4 & -22.21 & 30.64 & 10.42 & 9.04 & 8.7 & 0.84 \\
2002ke & 0.58 & 18.17 & -21.61 & -21.27 & 22.1 & 10.25 & 8.96 & 7.67 & 0.44 \\
2002kl & 0.41 & 5.91 & -19.07 & -18.9 & 0.6 & 8.86 & 8.3 & 7.35 & 0.14 \\
2003ba & 0.29 & 8.18 & -20.93 & -20.42 & 18.5 & 10.16 & 8.92 & 8.48 & 0.82 \\
2003bb & 0.96 & 20.37 & -23.3 & -22.77 & 173.35 & 11.3 & 9.46 & 7.97 & 0.18 \\
2003bc & 0.51 & 4.45 & -20.65 & -20.43 & 6.27 & 9.52 & 8.61 & 7.85 & 0.2 \\
2003dx & 0.51 & 2.17 & -19.19 & -18.94 & 1.59 & 9.15 & 8.43 & 8.34 & 0.45 \\
\textit{2003dz} & 0.48 & 2.47 & -16.88 & -16.73 & 0.53 & 8.65 & 8.2 & 7.64 & 0.61 \\
2003ea & 0.98 & 4.38 & -20.36 & -20.21 & 4.81 & 9.47 & 8.59 & 8.74 & 0.57 \\
\textit{2003en} & 0.54 & 1.64 & -17.39 & -17.19 & 0.14 & 8.03 & 7.9 & 8.61 & 0.91 \\
2003er & 0.63 & 7.16 & -22.11 & -21.68 & 32.74 & 10.73 & 9.19 & 8.02 & 0.08 \\
2003et & 1.3 & 4.97 & -21.63 & -21.51 & 48.3 & 10.56 & 9.11 & 8.62 & 0.86 \\
2003ew & 0.58 & 15.21 & -20.58 & -20.18 & 10.15 & 9.86 & 8.77 & 8.48 & 0.71 \\
2003N & 0.43 & 3.73 & -17.51 & -17.15 & 1.66 & 9.23 & 8.48 & 7.89 & 0.69 \\
K0404-005 & 0.79 & 8.34 & -22.29 & -21.66 & 21.52 & 10.94 & 9.29 & 8.81 & 0.61 \\
\textit{K0404-003} & 0.55 & 1.13 & -15.54 & -15.37 & 0.16 & 8.06 & 7.91 & 7.65 & 0.56 \\
K0404-006 & 0.41 & 2.4 & -18.31 & -18.02 & 2.94 & 9.52 & 8.61 & 8.48 & 0.79 \\
K0404-008 & 0.28 & 9.45 & -21.16 & -20.59 & 27.12 & 10.54 & 9.1 & 9.0 & 0.7 \\
\textit{K0404-010} & 0.61 & 2.31 & -18.83 & -18.08 & 0.17 & 9.09 & 8.41 & 8.4 & 0.59 \\
K0405-001 & 1.01 & 11.0 & -22.7 & -22.48 & 196.31 & 10.35 & 9.01 & 8.19 & 0.28 \\
K0405-002 & 0.56 & 8.43 & -21.18 & -20.9 & 5.63 & 9.92 & 8.8 & 8.46 & 0.8 \\
K0405-005 & 0.68 & 2.55 & -18.17 & -18.05 & 0.4 & 8.48 & 8.12 & 7.93 & 0.3 \\
K0405-007 & 0.5 & 4.78 & -19.73 & -19.28 & 1.72 & 9.46 & 8.58 & 9.32 & 0.98 \\
\textit{K0405-008} & 0.88 & 3.32 & -18.21 & -17.72 & 1.85 & 9.17 & 8.45 & 8.03 & 0.6 \\
HST04Pata & 0.41 & 9.53 & -21.87 & -21.47 & 33.12 & 10.46 & 9.06 & 8.6 & 0.53 \\
\textit{HST04Cli} & 0.75 & 1.52 & -17.45 & -17.33 & 0.85 & 8.89 & 8.31 & 8.23 & 0.72 \\
HST04Wil & 0.42 & 8.3 & -20.2 & -19.9 & 2.41 & 9.49 & 8.6 & 8.27 & 0.69 \\
HST04Pol & 0.56 & 7.9 & -21.47 & -21.14 & 14.89 & 10.3 & 8.99 & 7.87 & 0.14 \\
HST04Jef & 0.96 & 2.26 & -18.37 & -18.31 & 0.41 & 8.48 & 8.12 & 8.12 & 0.69 \\
HST04Ken & 0.52 & 5.28 & -20.53 & -20.13 & 2.34 & 9.75 & 8.72 & 8.38 & 0.7 \\
HST04Cum & 0.97 & 3.44 & -18.78 & -18.72 & 2.93 & 9.14 & 8.43 & 8.3 & 0.69 \\
\textit{HST04Cay} & 0.8 & 1.15 & -17.61 & -17.41 & 1.5 & 8.8 & 8.27 & 7.9 & 0.2 \\
HST04Bon & 0.66 & 8.49 & -22.15 & -21.57 & 71.59 & 10.85 & 9.25 & 8.09 & 0.19 \\
\textit{HST04Sos} & 0.55 & 4.41 & -20.13 & -19.83 & 4.13 & 9.66 & 8.68 & 8.46 & 0.8 \\
\textit{HST04Fox} & 0.69 & 2.33 & -18.59 & -18.49 & 0.56 & 8.64 & 8.19 & 8.07 & 0.35 \\
HST04Con & 0.84 & 7.62 & -21.27 & -20.97 & 9.99 & 10.13 & 8.91 & 8.23 & 0.5 \\
HST04Hei & 0.58 & 14.92 & -22.29 & -22.06 & 31.05 & 10.43 & 9.05 & 7.4 & 0.14 \\
HST04Riv & 0.61 & 2.42 & -17.43 & -17.27 & 0.35 & 8.38 & 8.07 & 7.99 & 0.58 \\
HST04Geo & 0.94 & 5.13 & -20.09 & -19.97 & 3.34 & 9.28 & 8.5 & 8.62 & 0.85 \\
\textit{HST04Gua} & 1.26 & 4.19 & -22.9 & -22.06 & 117.58 & 11.49 & 9.55 & 8.48 & 0.43 \\
\textit{HST04Ida} & 0.91 & 1.59 & -17.14 & -17.1 & 0.51 & 8.63 & 8.19 & 8.38 & 0.77 \\
\textit{HST05Kirk} & 0.45 & 2.65 & -17.49 & -17.36 & 2.37 & 7.7 & 7.74 & 8.13 & 0.74 \\
HST05Pic & 0.91 & 6.0 & -20.49 & -20.42 & 4.31 & 9.41 & 8.56 & 8.3 & 0.62 \\
HST05Sev & 0.96 & 7.61 & -19.87 & -19.87 & 1.67 & 8.94 & 8.34 & 7.6 & 0.07 \\
\textit{HST05Sco} & 0.93 & 3.5 & -18.96 & -18.79 & 3.79 & 9.65 & 8.68 & 7.56 & 0.0 \\
\textit{HST05Boy} & 0.66 & 2.28 & -17.45 & -17.47 & 0.26 & 8.0 & 7.89 & 8.24 & 0.69 \\
HST05Den & 0.97 & 3.09 & -19.82 & -19.67 & 2.97 & 9.46 & 8.59 & 8.53 & 0.87 \\
HST05Bra & 0.48 & 2.85 & -20.18 & -19.8 & 4.59 & 9.74 & 8.72 & 9.01 & 0.94 \\
HST05Str & 1.03 & 4.05 & -20.56 & -20.37 & 9.52 & 9.72 & 8.71 & 7.31 & 0.0 \\
\textit{HST05Cas} & 0.73 & 1.47 & -17.68 & -17.61 & 0.33 & 7.96 & 7.87 & 8.09 & 0.77 \\
HST05Mob & 0.68 & 4.25 & -19.79 & -19.47 & 4.86 & 9.71 & 8.7 & 8.1 & 0.32 \\
HST05Ton & 0.78 & 6.75 & -21.73 & -21.31 & 25.92 & 10.56 & 9.11 & 8.6 & 0.76 \\
\textit{HST05Fil} & 1.21 & 2.73 & -19.37 & -19.38 & 4.28 & 9.33 & 8.52 & 7.66 & 0.0 \\
HST05Ste & 0.47 & 7.1 & -18.37 & -18.27 & 0.6 & 8.41 & 8.08 & 7.7 & 0.88 \\

      \hline
    \end{tabular}
    \label{tab:CC}}
\end{table*}

\begin{table*}
  \centering
  \caption{As table ~\ref{tab:CC} but for GRB host galaxies. Surface luminosity and $F_{light}$ 
depend on accurate positional information, hence, they are only calculated for hosts with {\it HST} imaging and positional
errors $<0.1$ arcsec and $<0.15$ arcsec respectively. } 
  \begin{tabular}{l| l l l l l l l l l l}
    \hline
    \hline
 
    GRB name&  {\it z} &$\rm r^{80}$ &  $\rm M_V$&$\rm M_B$& $\rm SFR$ & $\rm \log M_{\star} $& $\rm 12$ & Surface Lum & $\rm F_{light}$\\
    &   &$\rm [kpc]$ &  AB mag& AB mag& $\rm [M_{\odot}/yr]$ & $\rm [M_{\odot}] $ &$\rm +\log(O/H) $ &${\rm [L_{\odot}kpc^{-2}]}$& \\
    \hline

GRB970228&0.695&3.2&-18.13&-18.04&0.25&8.21&7.99&   &   \\
GRB970508&0.835&1.48&-18.37&-18.22&3.08&8.24&8.0&8.48&1.0\\
GRB970828&0.958&2.8&-19.43&-18.8&2.17&9.57&8.64&   &   \\
GRB980326&1.0&   &-12.81&-13.24&0.01&4.71&6.31&   &1.0\\
GRB980425&0.0085&   &-18.34&-18.09&0.34&8.53&8.14&   &   \\
GRB980613&1.1&3.75&-20.77&-20.42&6.34&9.83&8.76&   &0.42\\
GRB980703&0.97&2.42&-21.49&-21.23&53.79&10.15&8.92&   &0.56\\
GRB990705&0.86&9.38&-19.57&-19.98&3.31&7.89&7.84&   &   \\
GRB990712&0.43&2.25&-19.57&-19.43&1.07&8.94&8.33&8.39&0.97\\
GRB991208&0.71&1.16&-18.8&-18.68&0.55&8.59&8.17&   &0.94\\
GRB991216&1.02&2.25&-15.94&-16.3&0.13&6.26&7.05&   &   \\
GRB000210&0.846&   &-20.01&-19.85&1.89&9.21&8.47&   &   \\
GRB000418&1.12&1.7&-20.55&-20.48&18.16&9.14&8.43&   &0.45\\
GRB000911&1.06&   &-19.37&-19.2&1.36&9.09&8.41&   &   \\
GRB010921&0.45&2.76&-20.17&-19.87&1.74&9.38&8.54&8.62&0.44\\
GRB011121&0.36&5.89&-20.14&-19.75&1.4&9.55&8.63&8.36&0.51\\
GRB020405&0.69&   &-21.06&-20.75&4.96&9.89&8.79&8.31&0.59\\
GRB020819&0.41&   &-22.06&-21.53&14.5&10.52&9.09&   &   \\
GRB020903&0.25&1.43&-19.33&-19.34&1.02&8.69&8.22&8.44&0.96\\
GRB021211&1.006&1.63&-19.95&-19.12&6.95&10.26&8.97&8.67&0.76\\
GRB030329&0.17&1.03&-16.67&-16.52&0.87&7.47&7.63&8.16&0.99\\
GRB031203&0.1055&   &-19.07&-18.52&0.44&9.24&8.48&   &   \\
GRB040924&0.859&3.23&-19.55&-19.1&4.54&9.36&8.54&   &   \\
GRB041006&0.716&5.19&-18.73&-18.29&1.17&9.69&8.69&8.23&   \\
GRB050223&0.5915&   &-20.77&-20.51&4.3&9.81&8.75&   &   \\
GRB050416A&0.6535&2.12&-18.96&-19.38&1.77&7.58&7.68&8.98&0.97\\
GRB050525A&0.606&1.76&-16.25&-16.68&0.15&6.31&7.08&8.19&0.95\\
GRB050824&0.83&   &-18.62&-19.02&1.37&7.45&7.62&   &   \\
GRB050826&0.296&   &-20.97&-20.28&1.39&9.93&8.81&   &   \\
GRB051016B&0.9364&   &-19.35&-19.77&2.54&7.76&7.77&   &   \\
GRB051022&0.807&   &-21.55&-21.23&23.85&10.49&9.07&   &   \\
GRB060218&0.0331&0.55&-15.92&-15.92&0.05&7.44&7.62&   &   \\
GRB061126&1.1588&   &-22.36&-21.61&51.34&11.16&9.4&   &   \\
GRB080319B&0.937&   &-17.49&-17.23&0.13&8.07&7.92&8.58&   \\

    \hline
  \end{tabular}
  \label{tab:GRB}
\end{table*}
 
\large
\begin{table}
  \centering
  \begin{tabular}{l| c c}
    \hline
    \hline 
    & $P_{KS}(all)$\\
    \hline
    $M_V$          & 0.41    \\
    $M_B$          & 0.39    \\
    $B-V$            & 0.23   \\
    $SFR$          & 0.15    \\
    $\Sigma$        &0.14    \\
    $M_{\star}$    & 0.12     \\
    $\Phi$         & 0.04    \\
    $L_{surface}$   & 0.01   \\
    $r_{80}$        & 0.003  \\

    $F_{light}$ & 5$\times 10^{-3}$ \\
    \hline
  \end{tabular}
  \caption{KS probabilities for comparison of physical properties between GRB and CCSN host galaxies. Showing
    the probabilities that the distributions of each parameter are drawn from the same population. The parameters
    compared are the global star formation rates (SFR), the absolute B and V band luminosities ($M_V$ and $M_B$),
    the B-V colour, the luminosity of the pixel underlying each GRB/SN $L_{surface}$, the 80\% light radii $r_{80}$,
    the specific star formation rate $\Phi$, the surface star formation rate $\Sigma$ and the location of the
    GRB/SN on their cumulative host galaxy light.  }
  \label{tab:Pks}
\end{table}

% ############################################################################################################
%
% ############################################################################################################

\section{Selection effects}
It is clear from the above results that there are differences between the two samples in several 
comparative properties (e.g. $r_{80}$, surface brightness), while others (e.g. absolute magnitudes) appear
broadly similar. A key question is therefore what selection effects could plausibly operate within the sample, 
and how these might impact our comparisons, could they force the two disparate distributions to look 
rather similar? Or alternatively, might they create apparent differences in similar underlying distributions? Below,
we describe our motivation for our sample definition, and consider several selection effects, and their impact on 
the observed distributions of different parameters. 

In the selection of our sample we have attempted to be as inclusive as possible, that is, including essentially
all of the GRB hosts with $z<1.2$ (and any available photometry) and all of the candidate CCSN hosts found within the GOODS fields. 
It is however necessary to explore how a number of selection effects could impact the bias of the samples, 
and how these would be affected if further (more restrictive) criteria were imposed. 
Below we discuss the effects of redshift, SN type and extinction on the samples.

\subsection{Dust obscuration}

The perhaps most serious bias affecting GRB/CCSN selected galaxies is that incurred by dust obscuration along the line of sight. 
The brightest GRB optical afterglow observed is roughly 20 magnitudes brighter than a typical CCSN 
\citep{2008arXiv0803.3215B,2008Natur.455..183R}, and GRB afterglows typically remain
brighter than their associated SN for several days. 
Although a deeply buried burst could be expected to suffer from large extinctions and non-detected or very faint optical
afterglows, \citep[so called `dark' bursts, see e.g.][]{2001A&A...369..373F,2002MNRAS.330..583L,2004ApJ...617L..21J,2006ApJ...647..471L,2007ApJ...669.1098R,2009arXiv0905.0001P}  
dust destruction by X-rays could still be effective enough to allow UV/Optical observations of the afterglow according to 
\cite{2001ApJ...563..597F}. However, \cite{2009ApJS..185..526F} suggests very convincingly that dark bursts {\it may not} be representative
of the general GRB population, and trace different environmental properties than bursts with detected optical afterglows.
Either way, even in the absence of any transient optical emission it is possible to identify a redshift for a GRB from its X-ray
identified host galaxy,  \citep[e.g. GRBs 970828 or 051022][]{1998ApJ...493L..27G,2007ApJ...669.1098R}. 
This relative insensitivity to dust obscuration is one of the key advantages of GRBs over many other techniques for high 
redshift exploration. Indeed, while it is interesting to note that both spiral host galaxies in the GRB sample
(GRB 990705 \citep{2000A&A...354..473M}) and GRB 020819 \citep{2004ApJ...617L..21J} ) are from bursts which were plausibly 
dust obscured, in general the GRB afterglow is much brighter than any SN, and hence if the low spiral fraction in
GRBs were due to dust obscuring many optical afterglows, we would expect to see an even stronger bias against
spiral galaxies in the CCSN sample, which is not the case.

Indeed, SN are likely much more strongly affected by dust that GRBs; studies of local starburst galaxies in the IR suggest
that a reasonable fraction of CCSN may occur in deeply enshrouded regions of their hosts \citep{2003A&A...401..519M}, essentially 
invisible to optical observations. This problem becomes even more extreme at moderate redshift, where optical 
observations probe rest-frame UV light, thus one may then suspect that the CCSN sample may be incomplete due to SN being
lost to dust extinction. Since the dustiest galaxies tend to be those which are most massive it is likely that any dust obscuration
would remove the brightest hosts from our sample, and would imply that any impact on a  CCSN selected galaxy population
 from dust, would most likely act to decrease its mass distribution.

Indeed, while MIPS observations of the GOODS fields \citep{2005ApJ...635.1022C} suggest
that $\sim$60\% of SN hosts are detected, this is not true for GRB host observations; \cite{2006ApJ...642..636L} find a detection
rate of only $\sim$ 20\% implying that
dust may well have a larger impact on CCSN detection than GRBs. In contradiction to this we note that the deeper observations 
of the CCSN host may be a factor in the higher detection rate, and that comparing the detection rate above a uniform depth results 
in more similar rates.

\subsection{Evolution of global properties}
Although both CCSN and GRBs originate from young systems, this does not necessarily indicate that the
relations between broad band properties and underlying physical conditions should be the same for each sample. 
Since we explicitly assume a direct proportionality between the K band an stellar mass, or U-band
and star formation rate, any systematic differences in these proportionalities between the two sample
could create a bias in the observed populations. The morphological properties of the CCSN hosts, 
combined with their redder colours suggest that there is a significant older population already in place. 
In a sense these galaxies should therefore be reasonably representative of the samples of local 
star forming galaxies from which the stellar-mass and star formation rate indicators are derived. 
In contrast, GRB hosts are apparently irregular, and several studies indicate they are extremely
young, with ages for the {\em dominant} stellar populations of under $10^7$ years  \citep[e.g.][]{2004A&A...425..913C,2009arXiv0907.4988L}
For very young systems
the K-band luminosity is dominated by young stars \citep[e.g.][]{2004A&A...418..913B}, and therefore may well be enhanced 
per unit stellar mass, such an effect would cause us to significantly {\em overestimate} the
GRB host galaxy masses. Secondly, in very young stellar systems ($t < 10^8$ years) the relation between
U-band luminosity and SFR is not constant, but {\em underestimates} the SFR for a given U-band luminosity
\citep{2007MNRAS.377.1024V}. In other words, the very young stellar ages derived from detailed
studies of individual GRB host systems \citep[e.g.][]{2009arXiv0907.4988L} suggest that our derived
properties for the GRB hosts may be systematically too massive, with too low a star formation rate. 
Were this corrected it is likely that the GRB and CCSN sample would seem more disparate than we observe. 
To partly quantify this effect it is relevant to note that not only is there a relationship between K-band luminosity
and stellar mass, but also between effective radius and stellar mass \citep{2003AJ....125.1849B,2009ApJ...695..101D}. 
Since the median sizes of the GRB and SN hosts differ by a factor a $\sim 2$, this would also suggest that
the median mass of a CCSN host would be a factor $\sim 4$ larger. In essence, it is not possible for both the GRB and SN hosts to satisfy both
of these relations, given the very young stellar ages of GRB hosts, and their likely impact on
the broadband properties we hence suggest that it is the morphological (and size) difference which
defines the GRB and SN populations, and that CCSN hosts are indeed typically more massive than 
those of GRBs.

\subsection{Redshift}

A further selection effect to consider is the origin of the redshifts for any given CCSN or GRB. For CCSN the broad-band
photometric data available enables the derivation of a photometric redshift (although see below). In contrast
most GRB hosts do not have this coverage and therefore redshifts come primarily from either
emission redshifts of the hosts or via absorption redshifts derived via observations of their 
afterglows. Although emission line flux is not directly proportional to host continuum magnitude
there is a broad dependence which means that emission line redshifts can normally only be derived 
from brighter hosts. In contrast absorption redshifts can be determined {\em independently} of
host magnitude \citep[e.g.][]{2002ApJ...581..981B,2003ApJ...597..699H,2004A&A...419..927V}, although this is not necessarily straightforward for
low redshift bursts where the UV metal lines are not redshifted into the optical band. The consequence
of this is that the requirement of a measured redshift biases our GRB sample toward intrinsically brighter hosts. Indeed,
if we perform a KS test between the hosts with absorption line spectra and those with emission line
redshifts we find that the sample with absorption redshifts is fainter than those with redshifts 
derived from emission lines; KS-probability of being drawn from the same distributions is only P$_{KS} = 0.001$. 
In other words, it is plausible (though not certain) that we are missing a population of intrinsically faint, low to moderate
redshift GRB hosts.

In part because of this above discussion we have included photometric redshifts for the CCSN sample where possible. Since,
if the photometry is sufficiently well sampled, they do provide a necessary handle on the faint hosts not observed with TKRS or GOODS/FORS2. 
Though exclusion of hosts without spectroscopic redshift, would narrow down the sources of random errors, it would also bias the sample
towards observationally bright, and thus, on average towards more luminous host galaxies. We note that the mean apparent
magnitudes and absolute magnitudes are 23.54 and -19.8 for the complete sample, and 22.79 and -20.5 for hosts with spectroscopic redshifts, 
hence we include all CCSN hosts in the sample, independently of how the redshift was determined.

\subsection{SN typing}

Approximately half of the CCSN are typed with low confidence (Bronze medal), hence there is a probability that
we have a fraction of SN Ia hosts in the sample. SN Ia can appear in both old stellar population due to long delay
times between star formation and explosion, as well as exploding rapidly after the formation of the progenitor system.
Since they are more likely than CCSN to occur in latent stellar populations, this could clearly affect the colours, star formation rates and 
specific star formation rates of the CCSN sample we have analysed. It is, however more difficult to 
determine how the mass distribution will be affected. Performing SED-fitting and estimating the host stellar masses
of the GOODS-detected SN Ia's gives a $\sim 0.2 dex$ higher mass distribution, though the KS-probability conclude they are 
consistent with a single distribution. 

As a further test to rule out that the results have been disturbed by mistyped SN, we perform the KS-test also on the sample containing 
only securely typed CCSN (Gold and Silver medal). 
We discover that the G+S sample are brighter in the V band absolute magnitudes, but not significantly more massive than the complete sample. 
Using this subsample the the absolute magnitudes are dissimilar to the GRB sample at a statistically significant level ($P_{KS} \sim 0.06$), 
and the mass distributions have almost unchanged $P_{KS} = 0.13$. 

However, we note that this in part may well be due to the reduced numbers of hosts
in the sample (G+S:23 , B:35) when culling by SN confidence level, as well as due the fact that this sample is also brighter in apparent magnitudes.  
We note that, though some influence cannot be ruled out, the conclusions are overall not changed by including or excluding parts 
of the sample based on SN typing.

While there is no evidence that SN Ic host galaxies differ from the hosts of other types if CCSN when considering global 
properties, \cite{2007arXiv0712.0430K} gives a strong indication that they typically lie on the brighter parts of the host.
We note that such a bias introduced by SN Ic in the sample would act to decrease the separation between the CCSN and GRB populations,
though this effect is most likely small and would only effect the $F_{light}$ and surface luminosity distributions, implying that 
their intrinsic distributions are even more separate.

\subsection{The overall impact of selection effects on the observed sample}
Above we have considered various biases which are likely to be operating within our sample of GRB and CCSN host galaxies. 
These include selection effects which are inevitably introduced into any magnitude/flux limited sample and also 
intrinsic systematic errors which propagate through our sample due to our incomplete knowledge of the detailed 
physical states of the galaxies we are studying. Overall, we consider the apparent differences in size and morphology to
be compelling. Although dust extinction will impact both SN and GRB hosts we believe it should impact SN more, and hence
the different morphologies observed are inconsistent it being a dominant selection effect. Similarly, the lack of GRB hosts with
photometric redshifts biases them to the brighter hosts, where emission line redshifts can be obtained, the difference between
apparent host luminosities of bursts with host emission, or afterglow absorption redshifts is indicative that there may be
a faint population of GRB hosts (currently GRBs without redshift measurements) omitted from our sample. Finally, the extreme
properties of the GRB stellar populations based on detailed population modelling \citep[e.g.][]{2008IAUS..255..162L} imply that
using empirically determined relationships between monochromatic luminosities and physical properties is not necessarily optimal. 
Hence we conclude that the environments of CCSN and GRBs are indeed different, and consider explanations for this below.

% ############################################################################################################
%
% ############################################################################################################

\section{Discussion}

Although supernovae and GRBs are closely related phenomena, one question of interest is the 
characteristic environments -- both local
and galactic -- in which they form. By contrasting the environments
of the two transient events we can obtain clues to their stellar progenitors. 
This in turn
provides observational constraints to the pathways which can create 
GRBs and is central to understanding any biases in using GRBs as
cosmological probes (e.g. as probes of star formation) as opposed to galaxy samples selected in
flux limited surveys. For example, our comparison with the MUSIC sample suggests that roughly a few percent of the
starformation tracked by CCSN and GRB is too faint to be included in the flux limited sample. Finally, the fraction of
stars which may create GRBs as a function of environmental properties 
can feed into predictions of high redshift (and hence low metallicity) GRB rates, as an input
for potential future GRB missions targeting high redshift GRBs (e.g. EXIST \footnote{http://exist.gsfc.nasa.gov/}).

The conclusion of F06 is echoed by our results, showing that GRB hosts are consistently fainter and have more irregular morphology than their 
SN counterparts. Given the well calibrated relation between luminosity
and metallicity, e.g. \cite{2004ApJ...613..898T}, this is most clearly explained by a preference for GRBs in low 
metallicity environments. 
F06 also compared how CCSN and GRBs trace blue light in the hosts. The findings are 
consistent with the CCSN tracing the blue light, and therefore broadly the global star-formation. 
The GRB population on the other hand appears to be significantly more concentrated on the brightest regions of  the 
galaxies.
This could naturally be interpreted as GRBs being due to the collapse of more massive stars, probably with 
initial 
masses $>$20 M$_{\odot}$ \citep{2007MNRAS.376.1285L}. These stars form in large OB-associations, and, since stellar 
luminosity traces a high power of 
stellar mass (crudely $L_{\star} \propto m_{star}^3$),
produce much more light than stars of lower mass, even those which produce supernovae. 

This is further reflected in an analysis of the surface brightnesses measured {\em directly} under
the transient position, which accepts the possibility that they are being drawn from the same
population with a KS-probability of only 0.01. Furthermore a comparison of locations within the hosts following the
method of F06 is even more compelling suggesting that the two distributions cannot
be reconciled with a probability higher than $P_{KS}=5 \times 10^{-3}$. These results are naturally explained by
the origin of GRBs in very young, and subsequently very massive stellar progenitors.

The so far most successful progenitor model for long GRBs is the collapsar model \citep{1993ApJ...405..273W}, 
predicting that the bursts are the result of the collapse of rapidly rotating  cores from 
massive stars. The metallicity to a large extent determines the rate of mass loss 
that is due to stellar wind in the progenitor star, and hence also the angular momentum
loss. Core collapse progenitors arising in low metallicity environments support only weak 
winds and may be able to retain a large fraction of the initial rotation.
As rapid rotation is thought to be one of the key the discriminators between GRB and CCSN explosions, it is natural to expect 
that GRB progenitors may therefore form in lower metallicity environments. However, all SN so far associated with GRBs are 
of the Ic variety, suggesting that the hydrogen
envelope has been lost, and indicating that simple low metallicity may not be sufficient to create
GRBs and that in single stars more exotic processes such as complete mixing on the main
sequence \citep[e.g.][]{2005A&A...443..643Y} may be necessary.

Introducing the option of a binary star evolution
(e.g. \cite{2006MNRAS.372.1351L}, \cite{2007Ap&SS.311..177V}, \cite{2004ApJ...607L..17P}) can
potentially create GRBs across a wider range of metallicity.  A binary scenario is suggested where two massive
($M> 8 \Mo$) stars after main sequence evolution and separation tightening through a common envelope
phase end up as a neutron star or black hole and helium core binary. Tidal locking of the helium cores rotation
enables enough angular momentum to create a torus, and the accretion of this onto the central compact object
at core collapse powers the GRB. 
Although this scenario remains possible at all metallicities, magnetic braking by a strong stellar wind 
could bias also binary progenitors towards low metallicity environments. 

The discrimination between
the different progenitor routes can potentially be made via metallicity measurement for the host
galaxies. While binary channels will operate at all metallicities (albeit with an increased rate
toward the lower end) single star evolution may produce a sharp cutoff in the metallicity 
at which GRBs can be created. The two possibilities can potentially be tested via metallicities for a large sample of GRB hosts. 

The task of host galaxy metallicity measurement is made difficult owing to the large redshift of many
bursts. Therefore, many studies of long burst host galaxy metallicities have used 
a luminosity-metallicity relation for the estimate. Other possibilities to measure the local
metallicity are by using the GRBs optical or X-ray afterglow as a probe, and study the absorption lines when it shines 
through the immediate environment, see for example 
\cite{2005A&A...442L..21S,2004A&A...419..927V,2005ApJ...634L..25C}.

\cite{2007MNRAS.375.1049W} studied the host metallicities using largely the same sample as 
F06, but with a more conservative redshift constraint. 
Their modelling of  metallicity dependent efficiency for producing GRBs suggests that progenitor
metallicity is of importance, their favoured model being one with constant efficiency up to nearly
solar composition and with a sharp cutoff, although they make the implicit assumption that
the shape of the mass metallicity relation for GRB hosts is the same as for field
galaxies. While this may be the case, it is far from clear \citep{2008AJ....135.1136M}.  
The authors also comment on the global versus local metallicity {\it within} the galaxy. Importantly, without
spatially resolved spectroscopy, the variations between metallicity in different parts of the galaxy
can be almost as large as the scatter in the M-Z relationship. Thus spectroscopy without spatial resolution
may not yield better results (for the progenitors metallicity) than using mass or K-band luminosity as proxy.

Our new sample of GRB and CCSN hosts is a factor of 2-4 larger than previously available 
samples, and with the broadband coverage allows us to derive physical parameters. It is interesting 
to investigate how our results may be interpreted in terms of the above discussion. 

In contrast to previous studies, we do not find highly significant (considering the KS-test) differences 
between the $\rm M_V$ or 
$\rm M_B$ distributions for GRB and CCSN hosts, although the median GRB hosts is roughly a factor of 2 fainter
 than the median of CCSN (see Figure ~\ref{fig:cdf1} where we plot the cumulative distribution function
of $M_v$). Considered alone, this is inconsistent with previous studies, although it should be noted
that the distinction in absolute magnitude is previous samples was the least significant of a number
of parameters compared. 
The origin of the apparent discrepancy between
our results and those of F06 is down to the combination of two factors.
Firstly, we attempt to derive absolute magnitudes based on spectral templates, rather
than assuming flat spectrum sources. Secondly, our larger sample of CCSN is apparently
fainter than the sample considered in F06. Indeed, the
mean apparent magnitude of the new CCSN sample is $\sim 1$ magnitude fainter, despite
a similar redshift distribution. Although the new larger sample of CCSN does not suggest an
overall globally different luminosity function it is particularly interesting to note that 
the sample of GRB hosts contain no galaxies brighter than $M_V \sim -22.4$, while the CCSN host
population continues to $M_V \sim -23.3$. Given the luminosity -- metallicity relations discussed above
this may well be consistent with a sharp cutoff in the metallicity at which a GRB can be 
created. Comparison of these two distributions with models for GRB efficiency in binary and
single star models as a function of metallicity may help to elucidate this further, although in
practise a still larger sample of GRB and CCSN hosts may be necessary to place 
strong constraints. 
The main bias bias effects on the distributions of B and V absolute magnitudes are redshift method, and dust 
obscured hosts. Both emission line redshifts and dust will bias the GRB sample towards brighter hosts, while
dust in CCSN hosts will give us a fainter sample - although a quantitative estimation of how large these effects are is difficult,
they are acting in opposite directions, suggesting a fainter true GRB host population and a brighter true CCSN population.

Since the absolute magnitude distributions of the two populations show only modest differences, it is unsurprising
that the global distributions of other parameters which depend directly on the magnitude in 
a given band (principally mass and star formation rate) are also similar. 
Further, since GRB hosts are on average bluer and of lower mass (even though the difference between
each distribution are not significant in their own right) the distinction in the specific star formation rate is much
stronger (this is also in part since the order of individual galaxies is obviously not identical in the
mass and SFR cumulative distributions). 
In Figure ~\ref{fig:SSFR}  we plot the specific star formation rates versus the stellar masses in the host galaxies. 
The majority of the GRB hosts are located in the low mass, high SSFR area, only a small fraction of the hosts
demonstrate high mass and low SSFR. The KS-test on the SSFR accepts, with a good statistical significance 
that GRB hosts typically have higher specific star formation rates than CCSN hosts. 

While the estimated stellar masses and starformation rates are compatible with a common distribution, we note that galaxy and 
stellar population age can have the effect on our measurements to overestimate the mass, and underestimate the SFR for 
young starbursts as discussed previously, while also dust obscuration will narrow  the mass distributions of the samples. 
Hence, it is {\em possible} that the mass and SFR distributions are more diverse than a  direct interpretation our results would indicate. 
This suggestion is further supported by simple morphological analysis of the host galaxy samples, 
which show striking differences. In the sample of CCSN hosts the
spiral fraction is approximately $\frac{27}{58} \sim 0.45$ with a Poisson counting error $\sim 5$. If the GRB 
host sample has identical spiral fraction, the expected number of spirals is $\sim 15 \pm 4$, whereas only two
can be recognised as spirals in the GRB host sample (GRBs 990705 and 020819)\footnote{This count ignores the unusual GRB 980425, 
but its inclusion only slightly affects the results}. The Poisson probability of two or less spiral galaxies to be found in a sample
with an expected spiral fraction of 0.45, is $\sim 4 \times 10^{-5}$. 

Performing a more quantitative analysis on the physical sizes of the
hosts reveals that GRB hosts are also significantly smaller than CCSN hosts. A comparison
of the 80\% light radii using the KS-test results in $P_{KS}=0.003$ that the sizes are drawn from the same parent distribution.
 In Figure ~\ref{fig:Msize} we plotted $\rm r_{80}$ versus $M_v$. Visual inspection confirms that the GRB host
population is smaller than the CCSN host population, which is accepted by the KS-test, and is in excellent agreement with 
with the morphological distribution - small irregulars versus large grand design spirals.

As an alternative to estimating mass from the K-band luminosity, we note that there is also a strong trend in the size-stellar mass
relation \citep[e.g.][]{2009MNRAS.396L..76S}. Since the luminosity based mass estimates suggest consistent 
distributions for the CCSN and GRB samples, but the size distributions are inconsistent, {\it both} of these relations cannot be correct.
Due to the uncertainties in stellar population ages, and their contributions to the K-band luminosities, we suggest that
size is a more stable proxy for mass when comparing samples of potentially different ages. 
Inserting the size distributions into any size-to-mass relation would hence yield a significantly lower mass
 distribution than estimated by the K-band luminosity and result in a KS-probability for the mass identical to that
of $r_{80}$. 
However, if this argument is wrong, and the K-band mass estimates are indeed correct, this would suggest that the
host masses are more similar than previously though, and implications on global environments and metallicities would
put constraints on the collapsar model.

The low probability of the size and morphological distributions being compatible is obviously in conflict with the apparently 
similar mass (K band luminosity) distributions discussed above, and does suggest markedly different large scale environments.
Assuming that GRB hosts have {\em similar mass} distributions but {\em smaller size} distribution than 
the CCSN host sample, we look at 
size - metallicity relations at constant mass; A positive correlation between 
size and metallicity is found by \cite{2007ApJS..173..441H} for UV selected and galaxies and by \cite{2008AJ....135.1877E} for galaxies 
in close pairs. On the opposite side, \cite{2008ApJ...672L.107E} indicate that the mass-metallicity relation in $\sim 44 000$ SDSS galaxies
is offset to higher metallicities for galaxies with decreasing size.

The ambiguity of these results can be interpreted in two ways: If the estimates mirror the true distributions, then we can deduce 
that GRB hosts, and progenitor stars, have similar mass and metallicity distributions, but have significantly higher stellar densities.
Alternatively, if the estimated mass distributions are dominated by galaxy-evolutionary or dust obscuration bias effects, then the 
GRB hosting population could be significantly less massive than it appears from the K-band estimates. Instead, if the mass-to-light ratio
is violated, galaxy size will be a more stable indicator if galaxy mass;  
This notion is supported by strong trends in the size-stellar mass relation \citep[e.g.][]{2009MNRAS.396L..76S}, which  also notes 
the age-dependency of this relation establishes smaller sizer for old galaxies at a given mass - hence we can be certain that
galaxy evolution is not a major concern for galactic sizes.

\section{Summary}

We have used multiwavelength photometry to investigate the physical 
properties of long gamma ray bursts and core-collapse supernovae 
hosting galaxies at low to intermediate redshifts. 
We fit spectral energy distributions, and estimate
restframe absolute magnitudes, stellar masses and star formation rates.
From the stellar masses we have also attempted to estimate host 
metallicities. Galaxy sizes and morphologies are studied.
Our results show that within our sample the derived masses and
absolute magnitudes are not significantly different between
the two populations, although the majority of likely selection effects
act to shrink any intrinsic separation within the two samples. Indeed,
while not statistically significant in terms of a KS test, the cutoff in the
luminosity function of GRB hosts about 1 magnitude fainter than the CCSN hosts, is suggestive of a metallicity cutoff. 
Further, the physical sizes and morphologies within the two samples
are different with high statistical significance, and this lends
further support to models in which GRBs form only in certain environmental 
conditions, most likely related to low mass and metallicity. 

Finally, the locations of the bursts and CCSN on their hosts, measured
both in absolute terms, and relative to their cumulative light distributions shows
GRBs to be highly concentrated on their host light, and to be occurring
in regions of high absolute surface brightness. 

To summarise our interpretation in terms of current models for GRB production we suggest the following
 
\begin{itemize}
\item GRB hosts are consistently smaller than CCSN hosts. 
\item The high surface brightness, surface star formation rates and relative locations on hosts 
suggest that GRBs are originating in a younger, and more massive stellar population.
\item This and other lines of evidence suggest that the dominant stellar populations in GRB hosts
are very young. This may introduce systematic errors which overestimate stellar mass and underestimate
star formation rates.

\end{itemize}

\section*{Acknowledgements}
KMS thank the University of Warwick for doctoral studentship, AJL and NRT thank STFC for postdoctoral and
senior fellowship awards.
This research has made use of the GHostS database (www.grbhosts.org), which is partly funded 
by \textit{Spitzer}/NASA grant RSA Agreement No. 1287913. Based on observations made with the NASA/ESA \textit{Hubble Space Telescope}, obtained from the data archive at the Space Telescope Science Institute. STScI is operated by the Association of Universities for Research in Astronomy, Inc. under NASA contract NAS 5-26555. The observations are associated
with programme numbers 11513, 10189, 9352, 9074, 9405, 10551, 10135, 9180 and 8688.
Finally, we also thank the anonymous referee for a constructive report.
\bibliographystyle{mn2e}
\bibliography{paper1}
\label{lastpage}

\newpage

\newpage

\appendix
\begin{center}
  {\bf APPENDIX}
\end{center}

\begin{table*}
\small{
  \centering
\caption{Photometric catalog over CCSN host galaxies in the GOODS 
fields. Errors are 1 sigma standard errors, limits are 3 sigma limiting magnitudes estimated 
from the sky background.} 
  \begin{tabular}{l|l l l l l l l }

    \hline 
    \hline
SN name & B & V & I & Z & J & H & K \\
    \hline
2002fv  &  28.94 $\pm$ 0.53  &  28.16 $\pm$ 0.21  &  26.78 $\pm$ 0.12  &  26.89 $\pm$ 0.17  &  $>$27.82  &  $>$24.17  &  $>$26.96  \\
2002fz  &  23.23 $\pm$ 0.19  &  22.4 $\pm$ 0.07  &  21.45 $\pm$ 0.07  &  21.11 $\pm$ 0.08  &  $>$23.78  &    &  20.01 $\pm$ 0.02  \\
2002hs  &  24.17 $\pm$ 0.17  &  23.93 $\pm$ 0.12  &  23.51 $\pm$ 0.18  &  23.06 $\pm$ 0.17  &  23.25 $\pm$ 0.05  &  23.02 $\pm$ 0.62  &  22.70 $\pm$ 0.05  \\
2002hq  &  21.93 $\pm$ 0.18  &  21.08 $\pm$ 0.06  &  20.19 $\pm$ 0.07  &  19.90 $\pm$ 0.08  &  19.45 $\pm$ 0.02  &  19.23 $\pm$ 0.14  &  18.85 $\pm$ 0.02  \\
2002kb  &  21.47 $\pm$ 0.14  &  20.64 $\pm$ 0.05  &  20 $\pm$ 0.07  &  19.78 $\pm$ 0.08  &  19.3 $\pm$ 0.02  &  19.18 $\pm$ 0.17  &  18.89 $\pm$ 0.03  \\
2002ke  &    &  21.45 $\pm$ 0.05  &  20.72 $\pm$ 0.07  &  20.47 $\pm$ 0.08  &    &    &    \\
2002kl  &  23.32 $\pm$ 0.13  &  22.69 $\pm$ 0.06  &  22.28 $\pm$ 0.1  &  22.18 $\pm$ 0.13  &    &    &    \\
2003ba  &  21.07 $\pm$ 0.04  &  20.06 $\pm$ 0.01  &  19.63 $\pm$ 0.02  &  19.43 $\pm$ 0.03  &    &    &    \\
2003bb  &  22.32 $\pm$ 0.29  &  21.62 $\pm$ 0.12  &  20.71 $\pm$ 0.13  &  20.24 $\pm$ 0.12  &    &    &    \\
2003bc  &  22.6 $\pm$ 0.05  &  21.78 $\pm$ 0.02  &  21.29 $\pm$ 0.03  &  21.14 $\pm$ 0.04  &    &    &    \\
2003dx  &  24.02 $\pm$ 0.04  &  23.31 $\pm$ 0.02  &  22.78 $\pm$ 0.02  &  22.65 $\pm$ 0.03  &    &    &    \\
2003dz  &  25.51 $\pm$ 0.18  &  25.28 $\pm$ 0.14  &  24.79 $\pm$ 0.19  &  24.57 $\pm$ 0.24  &    &    &    \\
2003en  &  25.78 $\pm$ 0.06  &  25.34 $\pm$ 0.04  &  24.53 $\pm$ 0.04  &  24.49 $\pm$ 0.04  &    &    &    \\
2003er  &  22.65 $\pm$ 0.12  &  21.40 $\pm$ 0.03  &  20.41 $\pm$ 0.03  &  20.05 $\pm$ 0.03  &    &    &    \\
2003et  &  23.34 $\pm$ 0.04  &  23.09 $\pm$ 0.03  &  22.73 $\pm$ 0.04  &  22.25 $\pm$ 0.04  &    &    &    \\
2003ew  &  23.55 $\pm$ 0.14  &  22.61 $\pm$ 0.05  &  21.76 $\pm$ 0.05  &  21.45 $\pm$ 0.06  &    &    &    \\
2003N  &  24.96 $\pm$ 0.16  &  24.7 $\pm$ 0.11  &  24.32 $\pm$ 0.17  &  23.88 $\pm$ 0.17  &    &    &    \\
K0404-005  &  24.95 $\pm$ 0.08  &  22.88 $\pm$ 0.01  &  21.23 $\pm$ 0.01  &  20.57 $\pm$ 0.0  &    &    &    \\
K0404-003  &  27.19 $\pm$ 0.14  &  27.13 $\pm$ 0.14  &  26.53 $\pm$ 0.16  &  26.43 $\pm$ 0.17  &    &    &    \\
K0404-006  &  24.03 $\pm$ 0.02  &  23.45 $\pm$ 0.01  &  23.02 $\pm$ 0.02  &  22.77 $\pm$ 0.02  &    &    &    \\
K0404-008  &  21.15 $\pm$ 0.01  &  19.84 $\pm$ 0.0  &  19.16 $\pm$ 0.0  &  18.83 $\pm$ 0.0  &    &    &    \\
K0404-010  &  27.45 $\pm$ 0.44  &  25.26 $\pm$ 0.05  &  23.76 $\pm$ 0.03  &  23.22 $\pm$ 0.02  &    &    &    \\
K0405-001  &  22.39 $\pm$ 0.01  &  21.66 $\pm$ 0.01  &  21.04 $\pm$ 0.01  &  20.87 $\pm$ 0.01  &    &    &    \\
K0405-002  &  22.39 $\pm$ 0.01  &  21.62 $\pm$ 0.01  &  21 $\pm$ 0.01  &  20.83 $\pm$ 0.01  &    &    &    \\
K0405-005  &  26.04 $\pm$ 0.11  &  25.24 $\pm$ 0.04  &  24.37 $\pm$ 0.04  &  24.33 $\pm$ 0.05  &    &    &    \\
K0405-007  &  24.14 $\pm$ 0.03  &  23.03 $\pm$ 0.01  &  22.21 $\pm$ 0.01  &  21.91 $\pm$ 0.01  &    &    &    \\
K0405-008  &  27.02 $\pm$ 0.23  &  26.22 $\pm$ 0.09  &  25.59 $\pm$ 0.1  &  24.89 $\pm$ 0.06  &    &    &    \\
HST04Pata  &    &  20.13 $\pm$ 0.0  &  19.56 $\pm$ 0.0  &  19.26 $\pm$ 0.0  &    &    &    \\
HST04Cli  &  26.92 $\pm$ 0.16  &  25.85 $\pm$ 0.05  &  25.42 $\pm$ 0.06  &  25.47 $\pm$ 0.08  &  24.22 $\pm$ 0.5  &    &  23.28 $\pm$ 0.32  \\
HST04Wil  &  22.65 $\pm$ 0.01  &  21.72 $\pm$ 0.01  &  21.27 $\pm$ 0.01  &  21.08 $\pm$ 0.01  &  20.83 $\pm$ 0.1  &  20.75 $\pm$ 0.1  &  20.61 $\pm$ 0.09  \\
HST04Pol  &  22.22 $\pm$ 0.01  &  21.43 $\pm$ 0.0  &  20.74 $\pm$ 0.0  &  20.5 $\pm$ 0.0  &  20.15 $\pm$ 0.07  &  19.91 $\pm$ 0.07  &  19.62 $\pm$ 0.06  \\
HST04Jef  &  25.7 $\pm$ 0.1  &  25.83 $\pm$ 0.1  &  24.99 $\pm$ 0.09  &  25.04 $\pm$ 0.13  &  $>$27.14  &  $>$23.63  &  $>$26.31  \\
HST04Ken  &  23.05 $\pm$ 0.02  &  22.21 $\pm$ 0.01  &  21.56 $\pm$ 0.01  &  $>$24.43  &  20.91 $\pm$ 0.1  &  20.74 $\pm$ 0.1  &  20.45 $\pm$ 0.08  \\
HST04Cum  &  25.17 $\pm$ 0.05  &  25.04 $\pm$ 0.04  &  24.58 $\pm$ 0.05  &  24.50 $\pm$ 0.05  &    &    &    \\
HST04Cay  &  26.75 $\pm$ 0.1  &  25.74 $\pm$ 0.03  &  25.58 $\pm$ 0.06  &  25.39 $\pm$ 0.06  &    &    &    \\
HST04Bon  &  23.56 $\pm$ 0.03  &  21.94 $\pm$ 0.01  &  20.67 $\pm$ 0.0  &  20.23 $\pm$ 0.0  &  19.59 $\pm$ 0.06  &  19.18 $\pm$ 0.05  &  18.81 $\pm$ 0.04  \\
HST04Sos  &  23.90 $\pm$ 0.03  &  22.8 $\pm$ 0.01  &  22.05 $\pm$ 0.01  &  21.76 $\pm$ 0.01  &  21.37 $\pm$ 0.13  &  21.22 $\pm$ 0.12  &  20.96 $\pm$ 0.11  \\
HST04Fox  &  24.91 $\pm$ 0.04  &  24.6 $\pm$ 0.02  &  24.01 $\pm$ 0.03  &  $>$26.36  &  23.92 $\pm$ 0.42  &  23.73 $\pm$ 0.39  &  23.43 $\pm$ 0.34  \\
HST04Con  &  23.43 $\pm$ 0.02  &  22.95 $\pm$ 0.01  &  22.08 $\pm$ 0.01  &  21.76 $\pm$ 0.01  &    &    &    \\
HST04Hei  &  21.47 $\pm$ 0.14  &  20.64 $\pm$ 0.05  &  20.00 $\pm$ 0.07  &  19.78 $\pm$ 0.08  &  19.3 $\pm$ 0.02  &  19.18 $\pm$ 0.17  &  18.89 $\pm$ 0.03  \\
HST04Riv  &  26.45 $\pm$ 0.13  &  25.64 $\pm$ 0.05  &  24.80 $\pm$ 0.05  &  $>$26.18  &  24.47 $\pm$ 0.56  &  25.18 $\pm$ 0.79  &  24.36 $\pm$ 0.53  \\
HST04Geo  &  24.26 $\pm$ 0.03  &  24.08 $\pm$ 0.03  &  23.36 $\pm$ 0.03  &  23.12 $\pm$ 0.02  &    &    &    \\
HST04Gua  &  26.11 $\pm$ 0.17  &  24.36 $\pm$ 0.04  &  22.66 $\pm$ 0.01  &  21.66 $\pm$ 0.01  &    &    &    \\
HST04Ida  &  27.10 $\pm$ 0.11  &  26.29 $\pm$ 0.08  &  26.49 $\pm$ 0.21  &  26.59 $\pm$ 0.3  &    &    &    \\
HST05Kir  &  24.66 $\pm$ 0.04  &  24.43 $\pm$ 0.03  &  23.98 $\pm$ 0.03  &  24.10 $\pm$ 0.05  &    &    &    \\
HST05Pic  &  23.60 $\pm$ 0.02  &  23.47 $\pm$ 0.02  &  22.81 $\pm$ 0.02  &  22.65 $\pm$ 0.02  &    &    &    \\
HST05Sev  &  24.15 $\pm$ 0.05  &  24.18 $\pm$ 0.04  &  23.66 $\pm$ 0.04  &  23.32 $\pm$ 0.04  &    &    &    \\
HST05Sco  &  25.20 $\pm$ 0.06  &  25.34 $\pm$ 0.06  &  24.58 $\pm$ 0.06  &  24.35 $\pm$ 0.06  &    &    &    \\
HST05Boy  &  25.45 $\pm$ 0.05  &  25.29 $\pm$ 0.04  &  24.80 $\pm$ 0.05  &  $>$26.37  &  24.29 $\pm$ 0.51  &    &  24.24 $\pm$ 0.52  \\
HST05Den  &  25.30 $\pm$ 0.07  &  24.78 $\pm$ 0.04  &  23.92 $\pm$ 0.03  &  23.53 $\pm$ 0.03  &    &    &    \\
HST05Bra  &  23.32 $\pm$ 0.02  &  22.28 $\pm$ 0.01  &  21.63 $\pm$ 0.01  &  21.36 $\pm$ 0.01  &    &    &    \\
HST05Str  &  24.03 $\pm$ 0.04  &  23.84 $\pm$ 0.04  &  23.21 $\pm$ 0.03  &  22.93 $\pm$ 0.03  &    &    &    \\
HST05Ste  &  24.34 $\pm$ 0.23  &  23.75 $\pm$ 0.09  &  23.32 $\pm$ 0.1   &  23.51 $\pm$ 0.1   &    &    &    \\
HST05Cas  &  26.33 $\pm$ 0.15  &  25.83 $\pm$ 0.08  &  24.98 $\pm$ 0.07  &  24.89 $\pm$ 0.08  &    &    &    \\
HST05Mob  &  24.91 $\pm$ 0.05  &  23.93 $\pm$ 0.02  &  22.97 $\pm$ 0.02  &  22.66 $\pm$ 0.01  &    &    &    \\
HST05Ton  &  23.22 $\pm$ 0.02  &  22.45 $\pm$ 0.01  &  21.45 $\pm$ 0.01  &  21.15 $\pm$ 0.01  &    &    &    \\
HST05Fil  &  24.94 $\pm$ 0.04  &  24.73 $\pm$ 0.03  &  24.57 $\pm$ 0.04  &  24.38 $\pm$ 0.04  &    &    &    \\
    \hline
  \end{tabular}
  \label{tab:phot}

}\end{table*}

\begin{table*}
\small{
  \centering
\caption{Photometric catalog continued: Spitzer IRAC bands}
  \begin{tabular}{l|l l l l} 
    \hline 
    \hline
SN name  & 3.6 $\mu m$ & 4.5 $\mu m$ & 5.8 $\mu m$ &  8 $\mu m$ \\
    \hline

2002fv  &   $>$ 25.65  &  24.52 $\pm$ 0.14  &   $>$ 23.58  &   $>$ 24.69  \\
2002fz  &     &     &     &     \\
2002hs  &  21.78 $\pm$ 0.01  &  21.79 $\pm$ 0.01  &  22.37 $\pm$ 0.06  &  22.53 $\pm$ 0.06  \\
2002hq  &  18.89 $\pm$ 0.01  &     &  19.39 $\pm$ 0.03  &     \\
2002kb  &  19.28 $\pm$ 0.01  &  19.95 $\pm$ 0.0  &  19.8 $\pm$ 0.13  &  19.74 $\pm$ 0.01  \\
2002ke  &  19.97 $\pm$ 0.01  &     &  20.47 $\pm$ 0.2  &     \\
2002kl  &  22.4 $\pm$ 0.03  &     &  23.17 $\pm$ 0.21  &     \\
2003ba  &     &  19.45 $\pm$ 0.01  &     &  18.55 $\pm$ 0.01  \\
2003bb  &     &  18.97 $\pm$ 0.01  &     &  19.43 $\pm$ 0.03  \\
2003bc  &     &     &     &     \\
2003dx  &     &  22.44 $\pm$ 0.02  &     &  22.46 $\pm$ 0.08  \\
2003dz  &     &  23.35 $\pm$ 0.04  &     &  23.66 $\pm$ 0.2  \\
2003ea  &  22.4 $\pm$ 0.07  &  22.78 $\pm$ 0.07  &  23.06 $\pm$ 0.38  &   $>$ 23.25  \\
2003en  &  24.58 $\pm$ 0.24  &  25.33 $\pm$ 0.25  &   $>$ 22.93  &   $>$ 25.44  \\
2003er  &     &  19.55 $\pm$ 0.0  &     &  20.08 $\pm$ 0.03  \\
2003et  &     &  20.89 $\pm$ 0.01  &     &  21.19 $\pm$ 0.02  \\
2003ew  &     &  21.16 $\pm$ 0.01  &     &  21.34 $\pm$ 0.05  \\
2003N  &  21.86 $\pm$ 0.02  &  21.86 $\pm$ 0.01  &  22.04 $\pm$ 0.07  &  22.2 $\pm$ 0.1  \\
K0404-005  &  18.99 $\pm$ 0.0  &  19.52 $\pm$ 0.0  &  19.67 $\pm$ 0.01  &  20.13 $\pm$ 0.04  \\
K0404-003  &     &  24.71 $\pm$ 0.16  &     &  23.44 $\pm$ 0.26  \\
K0404-006  &  21.03 $\pm$ 0.01  &     &  20.79 $\pm$ 0.02  &     \\
K0404-008  &  18.02 $\pm$ 0.0  &     &  18.33 $\pm$ 0.01  &     \\
K0404-010  &  21.71 $\pm$ 0.02  &  23.52 $\pm$ 0.03  &  22.76 $\pm$ 0.19  &  22.9 $\pm$ 0.12  \\
K0405-001  &     &  20.99 $\pm$ 0.01  &     &  21.19 $\pm$ 0.05  \\
K0405-002  &     &  20.98 $\pm$ 0.01  &     &  20.97 $\pm$ 0.05  \\
K0405-005  &  24.03 $\pm$ 0.07  &  24.29 $\pm$ 0.1  &   $>$ 23.97  &   $>$ 24.23  \\
K0405-007  &     &     &     &     \\
K0405-008  &  23.1 $\pm$ 0.06  &     &  23.15 $\pm$ 0.2  &     \\
HST04Pata  &  18.86 $\pm$ 0.0  &  19.26 $\pm$ 0.0  &  19.25 $\pm$ 0.03  &  17.97 $\pm$ 0.02  \\
HST04Cli  &  22.88 $\pm$ 0.13  &     &  22.47 $\pm$ 0.11  &     \\
HST04Wil  &  20.91 $\pm$ 0.03  &     &  21.43 $\pm$ 0.07  &     \\
HST04Pol  &  19.83 $\pm$ 0.01  &  20.2 $\pm$ 0.0  &  20.19 $\pm$ 0.04  &  20.28 $\pm$ 0.05  \\
HST04Jef  &     &     &     &     \\
HST04Ken  &     &     &     &     \\
HST04Cum  &  23.21 $\pm$ 0.05  &  23.5 $\pm$ 0.05  &  23.69 $\pm$ 0.23  &   $>$ 24.06  \\
HST04Cay  &  23.65 $\pm$ 0.05  &  23.83 $\pm$ 0.08  &  23.11 $\pm$ 0.27  &  23.79 $\pm$ 0.32  \\
HST04Bon  &  18.97 $\pm$ 0.0  &  19.32 $\pm$ 0.0  &  19.34 $\pm$ 0.01  &  19.51 $\pm$ 0.01  \\
HST04Sos  &  21.21 $\pm$ 0.02  &  21.54 $\pm$ 0.01  &  21.76 $\pm$ 0.08  &  21.82 $\pm$ 0.06  \\
HST04Fox  &     &  24.23 $\pm$ 0.09  &     &   $>$ 24.54  \\
HST04Con  &  20.77 $\pm$ 0.0  &  21.25 $\pm$ 0.01  &  21.2 $\pm$ 0.05  &  21.88 $\pm$ 0.07  \\
HST04Hei  &  19.28 $\pm$ 0.01  &  19.95 $\pm$ 0.0  &  19.8 $\pm$ 0.13  &  19.74 $\pm$ 0.01  \\
HST04Riv  &  23.89 $\pm$ 0.06  &     &  23.97 $\pm$ 0.44  &     \\
HST04Geo  &     &  23.13 $\pm$ 0.02  &     &  23.72 $\pm$ 0.23  \\
HST04Gua  &  18.74 $\pm$ 0.03  &     &  19.18 $\pm$ 0.04  &     \\
HST04Ida  &     &  24.32 $\pm$ 0.16  &     &   $>$ 24.21  \\
HST05Kir  &     &     &     &     \\
HST05Pic  &  22.42 $\pm$ 0.03  &     &  22.69 $\pm$ 0.26  &     \\
HST05Sev  &  23.51 $\pm$ 0.08  &     &  23.47 $\pm$ 0.38  &     \\
HST05Sco  &  22.06 $\pm$ 0.03  &     &  22.7 $\pm$ 0.17  &     \\
HST05Boy  &     &  24.8 $\pm$ 0.26  &     &   $>$ 24.47  \\
HST05Den  &  22.65 $\pm$ 0.03  &  22.75 $\pm$ 0.02  &  23.05 $\pm$ 0.12  &  23.05 $\pm$ 0.17  \\
HST05Bra  &  20.82 $\pm$ 0.0  &  21.04 $\pm$ 0.01  &  21.11 $\pm$ 0.03  &  20.93 $\pm$ 0.04  \\
HST05Str  &  21.99 $\pm$ 0.07  &  22.4 $\pm$ 0.06  &   $>$ 22.1  &  22.55 $\pm$ 0.13  \\
HST05Cas  &     &   $>$ 26.07  &     &   $>$ 24.67  \\
HST05Mob  &  21.3 $\pm$ 0.03  &  21.87 $\pm$ 0.01  &  21.85 $\pm$ 0.1  &  22.37 $\pm$ 0.09  \\
HST05Ton  &  19.79 $\pm$ 0.0  &  20.28 $\pm$ 0.0  &  20.23 $\pm$ 0.03  &  20.51 $\pm$ 0.02  \\
HST05Fil  &  23.32 $\pm$ 0.06  &     &   $>$ 24.25  &     \\
HST05Ste  &  23.54 $\pm$ 0.09  &     &   $>$ 23.81  &     \\

    \hline
  \end{tabular}
  \label{tab:phot2}

}\end{table*}

\begin{table*}
\small{
  \centering
\caption{GRB host photometry in the {\it Spitzer} IRAC bands. Limits are 3-sigma background estimates, errors are 1-sigma.}
  \begin{tabular}{l|l l l l } 
    \hline 
    \hline
GRB  & 3.6 $\mu m$ & 4.5 $\mu m$ & 5.8 $\mu m$ & 8 $\mu m$  \\
    \hline

970228  &  22.02  $\pm$  0.2  &  &$ >$ 20.02   &    \\
990712  &   &  21.98  $\pm$  0.4  & & $>$ 19.42   \\
991208  &   & $>$ 22.21   &   & $>$ 20.57   \\
000210  &   &  21.76  $\pm$  0.23  &   &  20.48  $\pm$  0.25  \\
000911  &   & $>$ 22.12  &  & $>$ 18.41  \\
010921  &  21.74  $\pm$  0.43  &  & $>$ 20.15   &    \\
020405  &  20.81  $\pm$  0.15  &   & $>$ 19.82           &    \\
020819  &  18.96  $\pm$  0.02  &   & 19.27 $\pm 0.22$   &    \\
021211  &    &  21.24  $\pm$  0.24  &  & $>$ 18.57   \\
030329  &  $>$22.59  & & $>$ 18.96   &   \\
031203  &  18.19  $\pm$  0.03  &  &  17.6  $\pm$  0.06  &   \\
040924  &  $>$21.92  & & $>$ 19.81   &  \\
041006  &  21.43  $\pm$  0.19  &  & $>$ 20.0    &  \\

    \hline
  \end{tabular}
  \label{tab:grb_irac}

}\end{table*}

\end{document}